\documentclass[twocolumn]{emulateapj}
\newcommand{\etal}{{~et al.~}}
\newcommand{\kms}{km~s$^{-1}$\,}

\newcommand{\msun}{{M$_\odot$}}
\newcommand{\mh}{M_{BH}}
\newcommand{\ms}{{\mh-\sigma}}

\newcommand{\beq}{\begin{equation}}
\newcommand{\eeq}{\end{equation}}

\newcommand{\lap}{\lesssim}

\def\Sec{\hbox{${}^{\prime\prime}$\llap{.}}}

\def\sec{\hbox{${}^{\prime\prime}$}}

\righthead{Nuclear Dynamics of NGC205}
\lefthead{Valluri \etal }

\begin{document}

\title{The Low End of the Supermassive Black Hole Mass Function: \\
Constraining the Mass of a Nuclear Black Hole in NGC 205 \\ 
via Stellar Kinematics}

\bigskip\bigskip 

\author{Monica Valluri,\altaffilmark{1}
Laura Ferrarese,\altaffilmark{2,3}
David Merritt,\altaffilmark{4} and
Charles L. Joseph\altaffilmark{2}
}
\altaffiltext{1}{Kavli Institute
for Cosmological Physics, University of Chicago, 5640 S. Ellis
Avenue, Chicago 60637\\
{\tt valluri@oddjob.uchicago.edu}
}
\altaffiltext{2}{Department of Physics and Astronomy,
Rutgers University, New Brunswick, NJ 08854\\
{cjoseph@physics.rutgers.edu}
}
\altaffiltext{3}{National Research Council Canada, Herzberg Institute 
of Astrophysics, Victoria, BC, V9E 2E7, Canada}

\altaffiltext{4}{Department of Physics, Rochester Institute of Technology, 
Rochester, NY 14623\\
{\tt merritt@astro.rit.edu}
}

\slugcomment{Accepted for Publication in The Astrophysical Journal, 20
July 2005, v628}


\begin{abstract}

Hubble Space Telescope (HST) images and spectra of the nucleated dwarf
elliptical galaxy NGC 205 are combined with 3-integral axisymmetric
dynamical models to constrain the mass $M_{BH}$ of a putative nuclear
black hole. This is only the second attempt, after M33, to use
resolved stellar kinematics to search for a nuclear black hole with
mass below $10^6$ solar masses.  We are unable to identify a best-fit
value of $M_{BH}$ in NGC 205; however, the data impose a upper limit
of $2.2\times 10^4$\msun~($1\sigma$ confidence) and and upper limit of
$3.8\times 10^4$\msun~($3\sigma$ confidence). This upper limit is
consistent with the extrapolation of the $M_{BH}-\sigma$ relation to
the $M_{BH}<10^6$\msun~ regime. If we {\it assume} that NGC 205 and
M33 both contain nuclear black holes, the upper limits on $M_{BH}$ in
the two galaxies imply a slope of $\sim 5.5$ or greater for the
$M_{BH}-\sigma$ relation.  We use our 3-integral models to evaluate
the relaxation time and stellar collision time in NGC 205; $T_r$ is
$\sim 10^8$ yr or less in the nucleus and $T_{coll}\approx 10^{11}$
yr. The low value of $T_r$ is consistent with core collapse having
already occurred, but we are unable to draw conclusions from nuclear
morphology about the presence or absence of a massive black hole.

\end{abstract}



\section{Introduction}
\label{sec:intro}

Evidence for the existence of supermassive black holes (SBHs) in
galactic centers has increased steadily in the past decade.
Kinematical detections with varying degrees of quality have been made
in roughly three dozen galaxies (see Ferrarese \& Ford 2004 for a
review), and virtually airtight evidence exists in two cases: our own
galactic center (Ghez et al. 2003; Sch\"{o}del et al. 2003) and NGC
4258 (Miyoshi et al. 1995).  The measured masses obey tantalizingly
tight relations with the bulge luminosity (Kormendy \& Richstone 1995;
McLure \& Dunlop 2002; Marconi \& Hunt 2003), the degree of
concentration of the bulge light (Graham et al. 2001; Erwin et
al. 2004), the central velocity dispersion of the stellar component
(Ferrarese \& Merritt 2000) and the velocity dispersion on kiloparsec
(Gebhardt et al. 2000a), and the virial velocity tens-of-kiloparsec
scales (Ferrarese 2002).

These relations are well established at masses exceeding $\sim
10^7$\msun, the regime spanned by all but two of the current
detections. Below $\sim 10^7$\msun~ only two measurements exist: in
the Milky Way ($\mh = 4\times10^6$ \msun, Ghez et al. 2003;
Sch\"{o}del et al. 2003) and in M32 ($\mh = 2.5\times10^6$ \msun,
Verolme et al. 2002, van der Marel et al. 1998).  Yet extending the
relations to the intermediate-mass black hole (IBH) regime ($\mh
\lesssim 10^6$\msun) is crucial for constraining models of black hole
formation (Haehnelt, Natarajan \& Rees 1998; Cattaneo, Haehnelt \&
Rees 1999; Monaco, Salucci \& Danese 2000; Bromley et al. 2004).  For
instance, the simple existence of IBHs would pose a serious challenge
for models in which nuclear black holes are born {\it in situ} from
the collapse of a ``supermassive star'' (e.g. Haehnelt et al. 1998). It would on the other hand, provide support for the
alternative scenarios in which SBHs evolve from the migration to
galactic centers of seed black holes produced, for instance, from the
collapse of Population III stars, or from dynamical processes in
ordinary star clusters (G\"urkan, Freitag \& Rasio 2004; Portegies Zwart et
al. 2004; Islam, Taylor \& Silk 2003, Portegies Zwart \& McMillan
2002; Miller \& Hamilton 2002).

Currently no fully compelling mass determinations exist in this
regime.  The interpretation of the super-luminous off-nuclear X-ray
sources (ULXs) detected in a number of nearby galaxies (Miller \&
Colbert 2004) is under debate.  Claimed kinematical detections of IBHs
in globular clusters (Gebhardt, Rich \& Ho 2002; Gerssen et al. 2002)
and dense star clusters (Maillard et al. 2004) have generally failed
to stand up to close scrutiny (Gerssen et al. 2003; Baumgardt et
al. 2003a\&b; McNamara, Harrison \& Anderson 2003). Although indirect
evidence exists for IBHs in Seyfert galaxies (Filippenko \& Ho 2003;
Greene \& Ho 2004), a clear dynamical detection is still lacking.  In
galactic nuclei, only one firm upper limit, of a few thousand solar
masses, has been established for an IBH, in the late-type spiral M33
(Merritt, Ferrarese \& Joseph 2001; Gebhardt et al. 2001).  There is a
simple reason for this rather unsatisfactory state of affairs: the
dynamical influence of a $\sim 10^5$\msun~IBH is dominant within $\sim
0.3$ pc (the so-called ``radius of influence''), assuming that the
observed scaling of nuclear properties with $\mh$ continues into the
IBH regime.  At optical wavelengths and at Hubble Space Telescope (HST)
resolution ($\sim$ 0\Sec1), this restricts the search to galaxies
closer than $\sim 1$ Mpc.  Several dE and dSph galaxies fit this
criterion, but the need to observe objects with high enough central
surface brightness leaves only one unexamined candidate: NGC 205, a
dEn companion to the Andromeda galaxy.  NGC 205 has been well studied
from the ground (e.g. Bica et al. 1990; Carter \& Sadler 1990; da
Costa \& Mould 1988; Lee 1996; Davidge 2003; and references
therein) and in imaging mode only, with HST (Jones et al. 1996;
Cappellari et al. 1999; Bertola et al. 1995). Along with M32 and the
Sagittarius Dwarf (Monaco et al. 2005) it is one of the three
spheroidals in the Local Group to boast a photometrically distinct
nucleus (van den Bergh 1999). The dominant stellar population in the
nucleus is known to have formed between 100 and 500 Myr ago, after
star formation had ceased in the central region (Lee 1996).
The galaxy shows no sign of nuclear activity (Ho, Filippenko \&
Sargent 2003; Fabbiano, Kim \& Trinchieri 1992; Condon et al. 1998).
Based on a measurement of the luminosity profile of the nucleus,
combined with an estimate of the nuclear M/L ratio, Jones et al.
(1996) estimate a mass for the nucleus of $9 \times 10^4$ \msun, which
therefore sets an upper limit to the mass of any putative nuclear IBH.
The relation between $\mh$ and the central stellar velocity dispersion
$\sigma$ predicts $\mh = 7.4\times 10^4$\msun~ (adopting a central
velocity dispersion of $\sigma =39 \pm 6$ \kms; Peterson \& Caldwell
1993).  The influence radius of such an IBH would be $\sim 0.2$ pc,
barely resolvable by HST at the galaxy's distance of 740~kpc
(Ferrarese et al. 2000).

In this paper we present new HST images and spectra of NGC 205
(\S~{\ref{sec:observations}); describe the modeling method and derive
estimates of the mass-to-light ratios for the inner and outer regions
(\S~{\ref{sec:modeling}); construct axisymmetric dynamical models that
are consistent with the kinematical data and search for a best-fit
value of $M_{BH}$ (\S~{\ref{sec:results}); discuss the implications
for black hole scaling relations (\S~{\ref{sec:scaling}}); and
discuss the constraints that the morphology of the nucleus place on
the existence of a massive black hole (\S~{\ref{sec:morphology}}).
\S~{\ref{sec:conclusions}} sums up.

\section{Observations and Data Analysis}
\label{sec:observations}

\subsection{HST/STIS Spectra}

The nucleus of NGC 205 was observed for a total of 24,132s with the
52\sec $\times$ 0\Sec1 slit of HST's Space Telescope Imaging
Spectrograph (STIS) between 15 October 2002, and 22 October 2002.  The
Ca II absorption triplet ($\lambda\lambda$ = 8498.06, 8542.14, 8662.17
\AA) was sampled with the G750M grating, covering the spectral region
between 8275 \AA~ and 8847 \AA~ with 19.8 \kms~pixel$^{-1}$ spectral
sampling (at 8500 \AA). The STIS CCD pixel scale is 0\Sec05, giving a
spatial resolution of 0\Sec115. Eleven spectra were each divided into
a pair of consecutive (``CR-split") exposures to facilitate removal of
cosmic rays; in addition, each spectrum was repeatedly shifted
relative to the first by 2.25, 4.5 and 6.75 pixels along the
dispersion direction, both to allow for correction of bad pixels and
to improve the spatial resolution.  The spectral calibration was
performed using the IRAF task CALSTIS following the standard procedure
outlined in the STIS Data Handbook (version 3.0, Mobasher, Corbin \&
Hsu 2001).

Correction for analog-to-digital conversion errors, bias and dark
subtraction, pixel-to-pixel flat field division, 2D rectification and
wavelength calibration were performed by the IRAF task CALSTIS, using
the most up-to-date calibration files provided by the Space Telescope
Science Institute. CALSTIS was also used to combine pairs of
``CR-split'' exposures, performing cosmic ray rejection in the
process, and to refine the wavelength calibration using spectral lamp
exposures obtained immediately before each spectrum. The latter step
is necessary to correct for spectral shifts due to positioning by the
Mode Select Mechanism and thermal motions.  In addition, each spectrum
was corrected for the ``fringe'' pattern which affects observations
longward of 7000 \AA, using flat fields obtained immediately before
each spectrum, each calibrated (using the same procedure adopted for
the NGC205 spectra) and normalized.  Finally, all spectra were
combined using the IRAF task DRIZZLE, with shifts calculated from the
astrometric information recorded in the ``jitter'' files.  The actual
recorded shifts between spectra were found to be within 0\Sec35 of the
commanded ones, well within the positioning uncertainties of the
instrument.

Spectra were extracted between $-$0\Sec266 and +0\Sec289 from the
nucleus and are shown in Figure~\ref{fig:spectra}.  At each location,
the line-of-sight velocity distribution (LOSVD) was determined by
deconvolving the nuclear spectra with one or more stellar templates,
using the maximum penalized likelihood algorithm (Merritt 1997).  Once
recovered nonparametrically, the LOSVDs were represented in terms of
Gauss-Hermite (GH) series (Gerhard 1993; van der Marel \& Franx 1993).
Confidence intervals on all quantities were computed via standard
bootstrap techniques.

\begin{figure}
\figurenum{1}
\epsscale{0.9}
\plotone{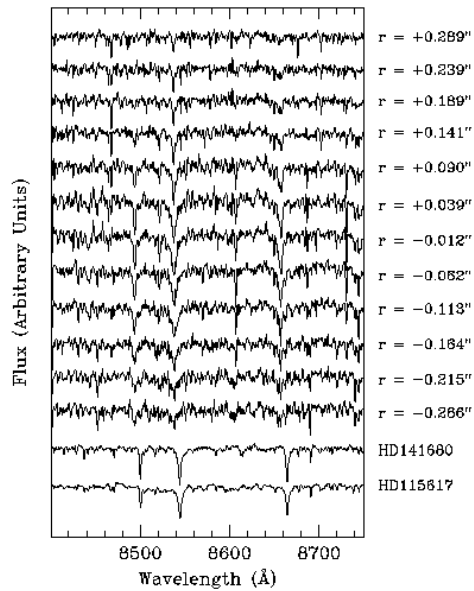}
\caption{STIS spectra obtained at 12 positions along the slit. The lowest
to frames show STIS spectra for the two template stars used in the analysis.}
\label{fig:spectra}
\end{figure}

Recovery of the LOSVDs requires knowledge of the spectrum of a star
representative of the entire stellar ensemble. In all the galaxies for
which STIS stellar kinematical measurements have been obtained from
stellar absorption lines (e.g. Gebhardt et al. 2003; Pinkney et
al. 2003; Cappellari et al. 2002), the stellar population in the
nucleus has been old enough that a single template star-- generally
that of the G8 III giant HD141680 -- has been adopted. Although other
template stars observed with the same instrumental configuration used
for the NGC 205 spectra exist in the HST archives, all but one are of
late G or early type K giant stars; the exception is HD115617, a main
sequence G5 star. As noted above, none of these standards is
representative of the stellar population in NGC~205, raising the
possibility that the kinematical parameters derived from the spectra
are biased as a consequence of template mismatch. To explore this
possibility, we have compared the LOSVDs derived using HD115617
(spectral type G8 III, hereafter Template 1) and HD141680 (spectral
type G5 V, hereafter Template 2) as templates: their spectra are
different enough (see Fig~\ref{fig:spectra}) to allow us to judge the
effects of template mismatch on the results.

\begin{table*}
  \caption{HST/STIS Kinematics for the Nucleus of NGC 205}
  \begin{tabular}{@{}c|rrrrrrrr|rrrrrrrr}
  \hline
{\footnotesize $r$} & \multicolumn{8}{c} {\footnotesize Template HD141680} & \multicolumn{8}{c} {\footnotesize Template HD115617}\\
    & {\footnotesize $V$} &  {\footnotesize $\Delta V$} &  {\footnotesize $\sigma$} &  {\footnotesize $\Delta \sigma$} & {\footnotesize $h_3$} & {\footnotesize $\Delta h_3$} &  {\footnotesize $h_4$} & {\footnotesize $\Delta h_4$} & {\footnotesize $V$} & {\footnotesize $\Delta V$} & {\footnotesize $\sigma$} & {\footnotesize $\Delta \sigma$} & {\footnotesize $h_3$} & {\footnotesize $\Delta h_3$} &  {\footnotesize $h_4$} & {\footnotesize $\Delta h_4$}  \\
{\footnotesize (1)} & {\footnotesize (2)}  & {\footnotesize (3} ) & {\footnotesize (4)}  & {\footnotesize (5)}  & {\footnotesize (6)}  & {\footnotesize (7)}  & {\footnotesize (8)} & {\footnotesize (9)}  & {\footnotesize (10)}  & {\footnotesize (11)}  & {\footnotesize (12)}  & {\footnotesize (13)}  & {\footnotesize (14)}  & {\footnotesize (15)}  & {\footnotesize (16)}  & {\footnotesize (17)} \tablenotemark{1}\\
\hline
{\footnotesize  -0.266} & {\footnotesize  4.62} & {\footnotesize 7.45} &{\footnotesize 17.04}  &{\footnotesize 5.65} &{\footnotesize -0.13} &{\footnotesize 0.13} &{\footnotesize 0.06} &  {\footnotesize0.05} & {\footnotesize 0.47}  & {\footnotesize  6.75}  & {\footnotesize 21.90}  &  {\footnotesize 6.20}  & {\footnotesize  0.01}  & {\footnotesize  0.07}  & {\footnotesize  0.00}  & {\footnotesize  0.04}\\
{\footnotesize  -0.215} & {\footnotesize  9.97} & {\footnotesize 5.80} &{\footnotesize 22.12}  &{\footnotesize 7.05} &{\footnotesize -0.14} &{\footnotesize 0.15} &{\footnotesize 0.05} &  {\footnotesize0.04} & {\footnotesize 13.23}  & {\footnotesize  6.60}  & {\footnotesize 28.16}  &  {\footnotesize 7.80}  & {\footnotesize  0.00}  & {\footnotesize  0.10}  & {\footnotesize -0.02}  &  {\footnotesize 0.06}\\
{\footnotesize  -0.164} & {\footnotesize -1.62} & {\footnotesize 4.30} &{\footnotesize 15.75}  &{\footnotesize 3.55} &{\footnotesize  0.05} &{\footnotesize 0.09} &{\footnotesize 0.05} &  {\footnotesize0.03} & {\footnotesize -1.09}  & {\footnotesize  3.95}  & {\footnotesize 17.59}  &  {\footnotesize 2.80}  & {\footnotesize  0.08}  & {\footnotesize  0.03}  & {\footnotesize  0.05}  &  {\footnotesize 0.02}\\
{\footnotesize  -0.113} & {\footnotesize -5.87} & {\footnotesize 4.00} &{\footnotesize 15.39}  &{\footnotesize 4.45} &{\footnotesize  0.03} &{\footnotesize 0.05} &{\footnotesize 0.06} &  {\footnotesize0.03} & {\footnotesize -5.02}  & {\footnotesize  3.65}  & {\footnotesize 16.87}  &  {\footnotesize 1.75}  & {\footnotesize  0.07}  & {\footnotesize  0.02}  & {\footnotesize  0.04}  &  {\footnotesize 0.01}\\
{\footnotesize  -0.062} & {\footnotesize -4.50} & {\footnotesize 4.05} &{\footnotesize 18.14}  &{\footnotesize 4.65} &{\footnotesize  0.00} &{\footnotesize 0.06} &{\footnotesize 0.09} &  {\footnotesize0.04} & {\footnotesize -5.57}  & {\footnotesize  3.85}  & {\footnotesize 19.46}  &  {\footnotesize 2.80}  & {\footnotesize  0.08}  & {\footnotesize  0.02}  & {\footnotesize  0.04}  &   {\footnotesize0.01}\\
{\footnotesize  -0.012} & {\footnotesize -0.40} & {\footnotesize 4.30} &{\footnotesize 20.55}  &{\footnotesize 4.00} &{\footnotesize  0.03} &{\footnotesize 0.07} &{\footnotesize 0.09} &  {\footnotesize0.03} & {\footnotesize -0.62}  & {\footnotesize  3.95}  & {\footnotesize 21.41}  &  {\footnotesize 3.35}  & {\footnotesize  0.05}  & {\footnotesize  0.06}  & {\footnotesize  0.03}  &  {\footnotesize 0.02}\\
{\footnotesize   0.039} & {\footnotesize  0.10} & {\footnotesize 3.60} &{\footnotesize 16.20}  &{\footnotesize 2.95} &{\footnotesize  0.06} &{\footnotesize 0.08} &{\footnotesize 0.04} &  {\footnotesize0.02} & {\footnotesize 2.85}  & {\footnotesize  2.80}  & {\footnotesize 15.93}  &  {\footnotesize 1.95}  & {\footnotesize  0.01}  & {\footnotesize  0.02}  & {\footnotesize  0.00}  & {\footnotesize  0.00}\\
{\footnotesize   0.090} & {\footnotesize -3.19} & {\footnotesize 4.00} &{\footnotesize 16.60}  &{\footnotesize 3.55} &{\footnotesize  0.05} &{\footnotesize 0.08} &{\footnotesize 0.06} &  {\footnotesize0.03} & {\footnotesize -0.12}  & {\footnotesize  3.70}  & {\footnotesize 18.52}  &  {\footnotesize 3.55}  & {\footnotesize  0.08}  & {\footnotesize  0.05}  & {\footnotesize  0.04}  &  {\footnotesize 0.03}\\
{\footnotesize   0.141} & {\footnotesize -4.12} & {\footnotesize 5.00} &{\footnotesize 17.69}  &{\footnotesize 4.85} &{\footnotesize  0.03} &{\footnotesize 0.07} &{\footnotesize 0.08} &  {\footnotesize0.04} & {\footnotesize 0.97}  & {\footnotesize  5.25}  & {\footnotesize 19.62}  &  {\footnotesize 4.55}  & {\footnotesize  0.06}  & {\footnotesize  0.06}  & {\footnotesize  0.04}  &  {\footnotesize 0.03}\\
{\footnotesize   0.189} & {\footnotesize -3.97} & {\footnotesize 4.95} &{\footnotesize 15.61}  &{\footnotesize 5.20} &{\footnotesize -0.10} &{\footnotesize 0.07} &{\footnotesize 0.06} &  {\footnotesize0.04} & {\footnotesize-0.72}  & {\footnotesize  6.35}  & {\footnotesize 19.60}  &  {\footnotesize 5.20}  & {\footnotesize -0.11}  & {\footnotesize  0.10}  & {\footnotesize  0.07}  &  {\footnotesize 0.06}\\
{\footnotesize   0.239} & {\footnotesize -0.19} & {\footnotesize 5.70} &{\footnotesize 11.73}  &{\footnotesize 2.80} &{\footnotesize -0.04} &{\footnotesize 0.04} &{\footnotesize 0.03} &  {\footnotesize0.02} & {\footnotesize 0.91}  & {\footnotesize  5.35}  & {\footnotesize 13.22}  &  {\footnotesize 2.05}  & {\footnotesize  0.00}  & {\footnotesize  0.03}  & {\footnotesize  0.01}  &  {\footnotesize 0.01}\\
{\footnotesize   0.289} & {\footnotesize -1.85} & {\footnotesize 6.00} &{\footnotesize 10.18}  &{\footnotesize 3.45} &{\footnotesize -0.02} &{\footnotesize 0.05} &{\footnotesize 0.02} &  {\footnotesize0.02} & {\footnotesize 1.64}  & {\footnotesize  5.05}  & {\footnotesize 11.57}  &  {\footnotesize 1.35}  & {\footnotesize  0.00}  & {\footnotesize  0.01}  & {\footnotesize  0.01}  &  {\footnotesize 0.00}\\
\hline
\tablenotetext{1}{Columns are: (1) Aperture distance from the 
morphological center of NGC 205;
(2, 10)  velocity (in \kms) relative to the heliocentric velocity of NGC205 (-241\kms); 
(3, 11) 1$\sigma$ error on $V$; 
(4, 12)  velocity dispersion (in \kms);
(5, 13) 1$\sigma$ error on $\sigma$; 
(6, 14)  Gauss-Hermite coefficient $h_3$;
(7, 15) 1$\sigma$ error on $h_3$; 
(8, 16) Gauss-Hermite coefficient $h_4$; 
(9, 17) 1$\sigma$ error on $h_4$.}
\end{tabular}
\end{table*}

The first four terms from the Gauss-Hermite expansion, $V$, $\sigma$,
$h_3$, and $h_4$, (van der Marel \& Franx 1993; Joseph et al. 2001)
derived from Template 1 and Template 2 are tabulated in Table 1 and
shown in Figure~\ref{fig:STIS-kin} as crosses and filled circles
respectively (the points from Template 2 have been offset slightly
along the horizontal axis to allow for better visual comparison).
Although the use of Template 1 resulted in a slightly better fit to
the NGC205 spectrum (as measured via the integrated square error), the
$V$ and $\sigma$ values from both templates are identical to within
$1\sigma$ statistical errors.  The overall consistency in the $h_3$
and $h_4$ terms is also quite good:  all values  are consistent with
zero.

\begin{figure*}
\figurenum{2}
\epsscale{0.9}
\plotone{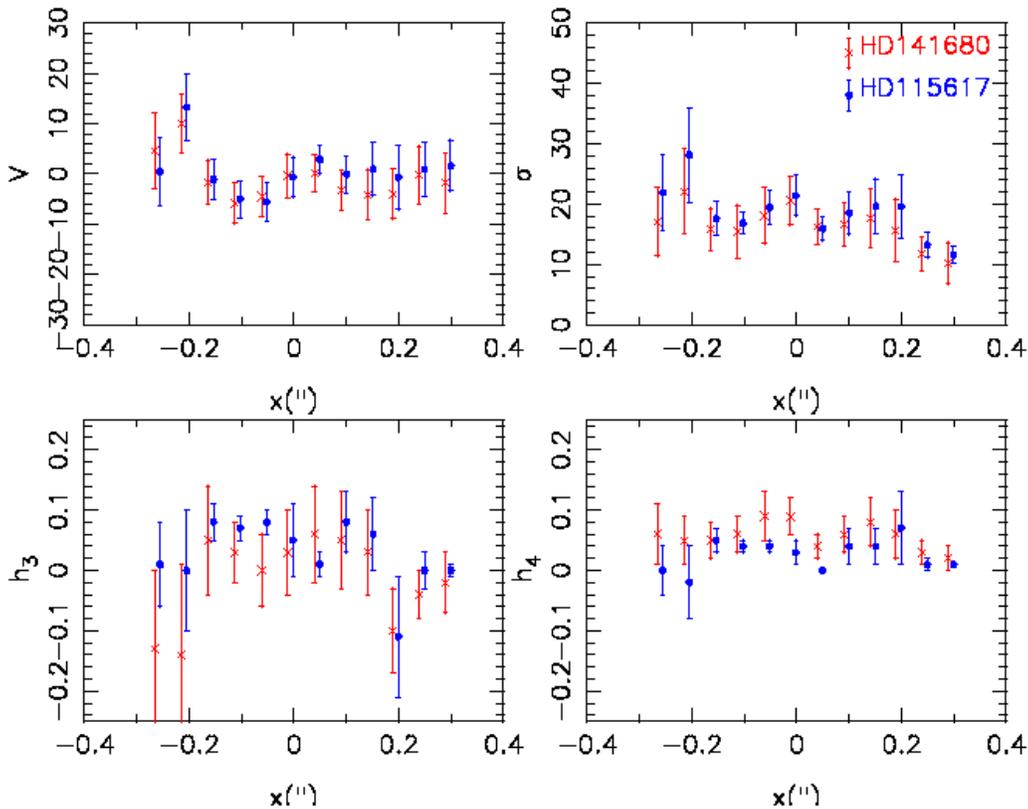}
\caption{ Nuclear kinematical data ($V, \sigma, h_3, h_4$) along the
slit obtained by deconvolving STIS spectra (Fig 1) with two different
stellar templates HD141680 (crosses) HD115617 (filled circles - which
have been offset slightly along the horizontal axis by an arbitrary amount
to enhance visibility.)}
\label{fig:STIS-kin}
\end{figure*}

The largest discrepancy is seen in the shape of the central ($r =
-0\Sec 012$) LOSVD, shown in Figure~\ref{fig:Central-LOSVD}; this is
particularly relevant since it is on the central LOSVD that a black
hole (if present) is expected to have the strongest effect.  The
noticable asymmetry in the central LOSVD obtained when using Template
1 (solid line) is absent when Template 2 is used instead (dot-dash
line). In an axisymmetric galaxy, strong asymmetries in the central
LOSVD are unphysical, and the asymmetry recovered with Template 1 is
therefore very likely the spurious result of template mismatch.

In summary, the consistency in the $V$, $\sigma$, $h_3$, and $h_4$
parameters derived using the two different templates gives us some
degree of confidence that template mismatch is unlikely to
significantly affect our results, although we should treat results
which make use of the full central LOSVD with caution. We further note
that  for stellar types earlier than F stars, temperature and rotation
increase the intrinsic width of the lines, and the presence of
hydrogen absorption lines from the Paschen series can further
complicate matters as they can overlap with the calcium features.  In
general, using an old stellar template to measure the kinematics of a
young stellar population in this spectral region will cause the
velocity dispersion, and therefore $\mh$, to be {\it
overestimated}.  Since only an upper limit to $\mh$ will be derived in
the following sections, we conclude that the upper limit could only
{\it decrease} if the correct stellar template were used. In other
words, the value derived below should be considered as a very firm upper
limit to the mass of a central IBH in NGC 205.

\begin{figure}
\figurenum{3}
\epsscale{0.75}
\plotone{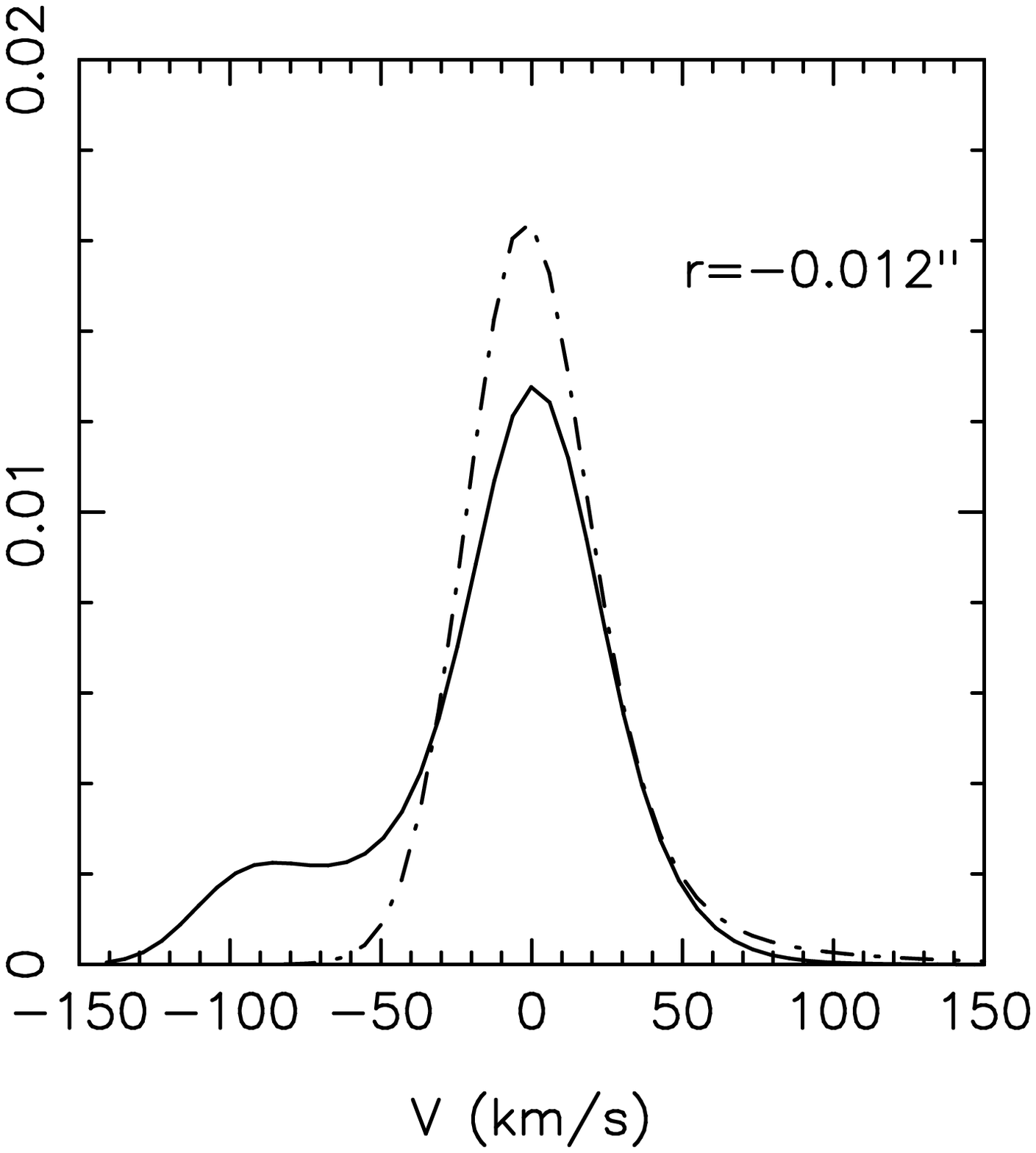}
\caption{LOSVD in STIS aperture closest to the center
($r=-0\Sec012$) obtained with the two different stellar
templates. Solid curve corresponds to deconvolution with Template 1 (a
G8 III star) and dot-dash curve corresponds to deconvolution with
Template 2 (a G5 III star).  The velocity is with respect to the
heliocentric velocity of the galaxy. The vertical axis is in arbitrary
units.}
\label{fig:Central-LOSVD}
\end{figure}

Finally, STIS data were complemented at large radii with the
ground-based kinematic data from Bender, Paquet \& Nieto (1991),
obtained with the 3.5m telescope at Calar Alto, consisting of 31
apertures within 84.56\sec.

\subsection{HST/ACS Images}

In order to determine the luminosity distribution in NGC 205, the
galaxy was observed on 03 October 2002 for a total of 2,440s with the
F814W filter ($\sim$ Johnsons' $I$) of the Advanced Camera for Surveys
(ACS) High Resolution Channel (HRC). The HRC pixel scale is 0\Sec027,
giving a field of view of 29\sec$\times$26\sec.  To improve the
spatial resolution, NGC 205 was observed at four different pointings,
each consisting of two CR-split images, arranged at the corners of a
square of side 0\Sec0989.  Total exposure times were 2,560s in F555W
and 2,440s in F814W.  Image calibration was performed using the IRAF
tasks CALACS and PYDRIZZLE following the standard procedure described
in the ACS Data Handbook (Mobasher, Corbin, \& Hsu 2003).  Basic
reduction (bias and dark subtraction, removal of overscan regions,
flat fielding and cosmic ray removal) was performed using IRAF's
CALACS, while the task PYDRIZZLE was used to correct for geometric
distortion, and combine the four drizzled images.

The structural parameters of NGC205 will be discussed in detail in a
forthcoming paper. Briefly, two independent procedures were adopted to
recover the isophotal parameters.  The IRAF task ELLIPSE (Jedrzejewski
1987) was employed to find the best fitting isophotes by iteratively
sampling the image along elliptical paths with given semi-major axis
length, keeping as free parameters the position of the center,
ellipticity, semi-major axis position angle, and high order
coefficients describing deviations of the isophotes from ellipses.
ELLIPSE is best applied to galaxies with a smooth surface brightness
distribution; in the case of NGC 205, the task fails to converge
beyond 0\Sec8, where the stellar population is clearly resolved.
Fischer et al. (1992) introduced a different approach, in which the
image is divided  in ellipsoidal annuli, each further divided into
eight sectors. The  median of the average brightnesses determined in
each sector is taken as the average brightness of each annulus. In the
case of NGC 205, the center and ellipticity of each annulus was set
equal to the  center found by ELLIPSE within 0\Sec8, and the average
ellipticity measured by Kim \& Lee (1998) and Lee (1996) from ground
based data respectively.
 
\begin{table}
  \caption{Composite $I-$band Surface Brightness Profile for NGC 205}
  \begin{tabular}{@{}rrrrrrr}
  \hline
$r$ &  $I$ & $\epsilon$ &&  $r$ &  $I$ & $\epsilon$ \\
(1) & (2) & (3) && (1) & (2) & (3)\tablenotemark{1}\\
 \hline
       0.025  &     12.35   &    0.2286  &  &         17.53   &    18.62  &     0.52\\
        0.05   &     12.48   &    0.234    &  &         22.91   &    18.7    &     0.29\\
       0.075   &    12.65   &    0.1953  &  &        24.29    &   18.8     &    0.27 \\
         0.1     &    12.85   &    0.0988  &  &       27.09     &  18.82    &    0.29  \\
       0.125   &    13.07   &    0.0082  &  &         30.01   &    18.86  &      0.3\\
        0.15    &    13.24   &    0.0566  &  &         33.01   &    18.94  &      0.31\\
       0.175   &    13.44   &    0.0891  &  &         37.12   &    18.99  &      0.33\\
         0.2     &    13.62   &    0.1132  &  &         41.45    &   19.06   &     0.34\\
       0.225   &    13.76   &    0.1453  &  &        44.91     &  19.11    &    0.35 \\
        0.25    &    13.91   &    0.1449  &  &         50.16    &   19.17   &     0.34\\
       0.275   &    14.09   &    0.1177  &  &          55.6      &  19.25    &    0.36\\
         0.3     &    14.31   &    0.0909  &  &         61.16     &  19.34    &    0.36\\
       0.325   &    14.51   &    0.0718  &  &         68.35     &  19.42    &    0.38\\
        0.35    &    14.67   &    0.0234  &  &         76.43     &  19.52    &    0.4\\
      0.3855  &    14.85   &    0.091    &  &         85.51     &  19.62    &    0.42\\
      0.5328  &    15.5     &    0.091    &  &        95.72      & 19.72     &   0.43\\
      0.7418  &    16.24   &    0.091    &  &        106.3      & 19.81     &   0.45\\
       1.037   &    16.82   &    0.091    &  &        118.0      & 19.9        &   0.46\\
       1.454   &    17.47   &    0.091    &  &        129.8     &  19.99      &   0.46\\
       2.043   &    17.76   &    0.091    &  &        144.1     &  20.1        &   0.47\\
       2.876   &    17.99   &    0.091    &  &        158.5     &  20.21      &   0.47\\
       4.051   &    18.24   &    0.091    &  &        179.5     &  20.33      &   0.5\\
       5.605   &    18.36   &    0.091    &  &        197.4     &  20.46      &   0.5\\
       7.543   &    18.43   &    0.091    &  &        217.2     &  20.61      &   0.5\\
       9.157   &    18.51   &    0.091    &  &        238.9     &  20.78      &   0.5\\
       13.92   &    18.56   &    0.3        &  &        262.8     &  20.95      &   0.5\\
\hline
\tablenotetext{1}{Columns are: (1) Semi-major axis length (in arcsec);
(2)  $I-$band magnitude; (3) Ellipticity. Data for $r \leq$ 0\Sec35 are derived
by applying IRAF/ELLIPSE to the ACS/HRC data. Data for 
0\Sec3855 $\leq r \leq$ 9\Sec157 are measured using the radial annuli technique
described in \S~\ref{sec:observations}, again from the HRC images. Data at larger
radii are based on ground-based data from Kim \& Lee (1998)}
\end{tabular}
\end{table}

The photometric calibration was performed using the zero points from
Sirianni et al. (2004) and a foreground extinction correction
$E(B-V)=0.081$ (Schlegel, Finkbeiner \& Davis 1998), giving
$A(I)=0.151$ for $R_V=A(V)/E(B-V)=3.1$ and $A(I)/A(V)=0.594$
(Cardelli, Clayton \& Mathis 1989).  The resulting surface brightness
profile is shown in Figure~\ref{fig:profile}, where we also plot the
$I-$band surface brightness profiles measured from ground-based data
by Kim \& Lee (1998) and Lee (1996). The ACS surface brightness
profiles determined using the two methods described above are in
excellent agreement. Both agree with the ground-based profiles except
in the innermost $\sim$ 1\sec, where the latter are strongly affected
by seeing.  The final, combined, composite $I-$band surface brightess
profile is tabulated in Table 2.


\begin{figure}
\figurenum{4}
\epsscale{0.95}
\plotone{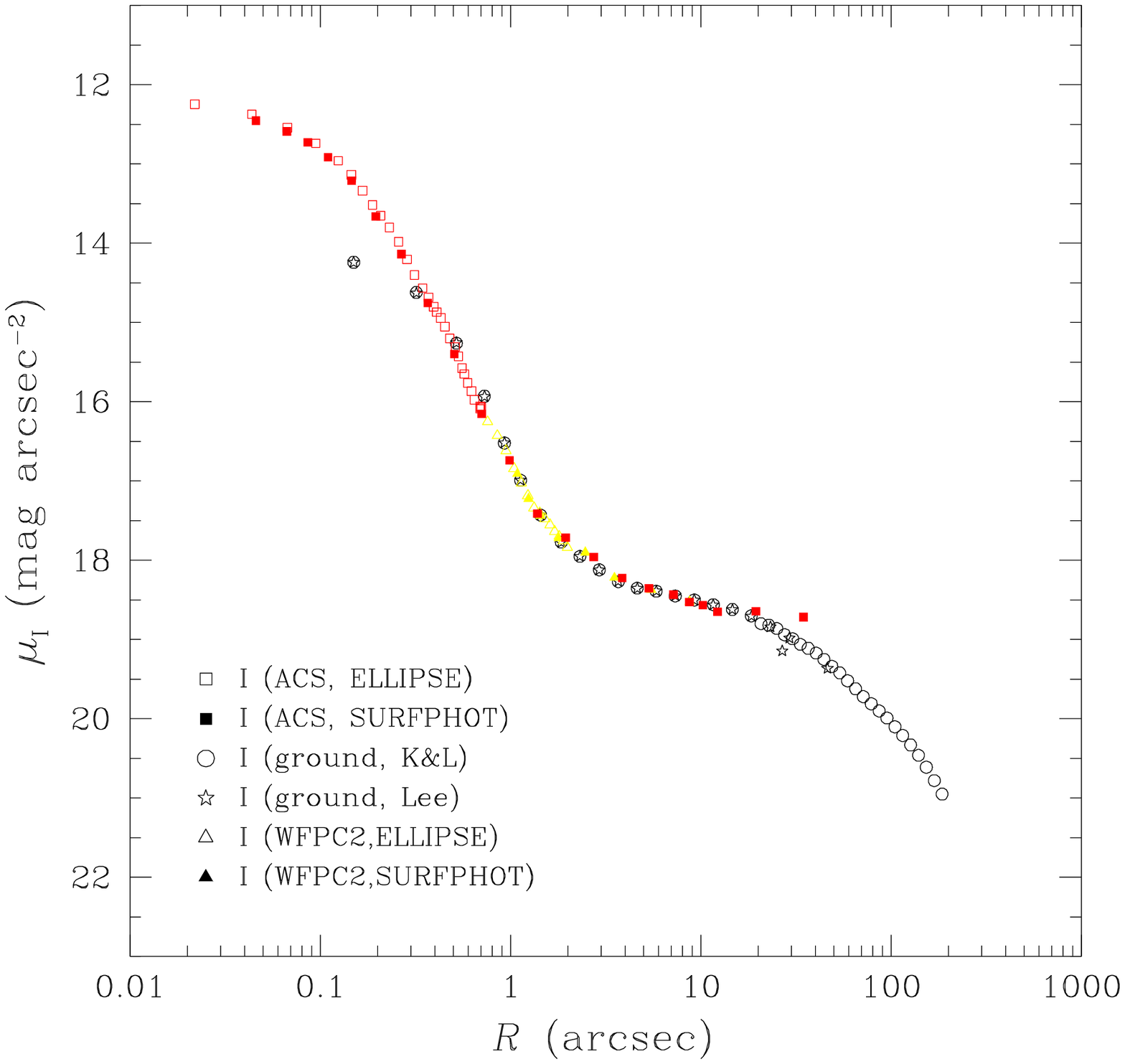}
\caption{The $I-$band surface brightness profile of NGC 205, obtained
by combining the HST/ACS observation discussed in this paper to the
ground-based data of Kim \& Lee (1998) and Lee (1996). The radius is
the ``effective radius'', given by $r = r_{SMA}\sqrt{1-\epsilon(r)}$,
where $r_{SMA}$ is the galaxy semi-major axis, and $\epsilon$ the
ellipticity.}
\label{fig:profile}
\end{figure}

The 3-dimensional luminosity density $j_*(\varpi,z)$ was computed via
a non-parametric algorithm (Merritt, Meylan \& Mayor 1997) under the
assumption that the galaxy is seen edge-on. The inner radius of the
surface brightness distribution 0\Sec025 and the outer radius of the
surface brightness distribution is 100\sec. The deprojection is
performed on a 2D logarithmically spaced grid with 100 grid points in
each dimension. The algorithm accounts for the varying ellipticity as
a function of radius and computes $j_*(\varpi,z)$ on the 2D grid by
minimizing the residuals between the projected $j_*$ and the observed
surface brightness. Since the deprojection is an unstable process,
smoothness was enforced via a ``thin-plate smoothing spline'' penalty
function (Wahba 1990).

\section{Dynamical Modeling}
\label{sec:modeling}

Dynamical constraints on the possible values of the parameters, the
black hole mass $\mh$ and the stellar M/L ratio $\Upsilon$, were
computed via orbital superposition (Schwarzschild 1979) by
constructing oblate spheroidal models for a large number of
[$\mh,\Upsilon$]. The observable properties of the models were then
compared to the surface brightness and spectral data.  With the
exceptions detailed in \S 3.1 and \S 3.2, the modeling algorithm used
is the one described in Valluri et al. (2004, hereafter VME04), to
which paper we refer the reader for a detailed description of the
method.  The numerically-computed orbits were ``observed'' at each of
the 31 ground-based apertures and the 12 STIS apertures.  For the STIS
apertures, a Gaussian PSF with FWHM of 0\Sec1 was assumed (Bower et
al. 2001), and the model results degraded accordingly.  Detailed
seeing information is not available for the ground-based data,
therefore the corresponding model results were not PSF-convolved. We
have found that the omission is unimportant as long as the data were
obtained with apertures wider than the seeing disk, which is indeed
the case for the ground-based observations used in the analysis.

A non-negative quadratic programming routine (E04NCF of the NAG
libraries) was used to find the weighted superposition of the orbits
that best reproduces both the assumed model stellar density
distribution $\rho(\varpi,z)$ and the observed kinematical data.  The
total number of luminosity constraints in NGC 205 is 304 (of which the
inner 128 are from the ACS photometry); kinematical constraints
consist of 62 contraints from the ground-based data (velocities $V$
and velocity dispersions $\sigma$ within 84\Sec56) and 40 contraints
from the STIS data ($V$, $\sigma$ and the GH moments $h_3, h_4$)
(Note: the last two apertures on the right end of the slit in Fig 2
were accidentally excluded from the dataset).

Since it is necessary to explore a wide range of the parameters
values: $\mh$, M/L-ratio (typically between 150-200 pairs of
parameters), we run an intitial set of models using a total of 4500
orbits per model.  Since the solution space is sensitive to the size
of the orbit library (see VME04), we follow these initial runs with a
limited number of models (keeping either $\mh$ or M/L-ratio fixed)
with larger libraries of 8100 orbits each. The CPU time required for a
single model with 8100 orbits is 7.5 hours on a 2.4GHz Intel Xeon
processor makes it prohibitively expensive to run a large number of
models at this resolution).  In the next two subsections we discuss two
aspects in which our models differ from those described in VME04.

\subsection{Radially Varying Mass to Light Ratios}
\label{sec:mbylprofile}

In the standard treatment of the 3-integral modeling problem the
luminosity density - derived as described in \S2.2 - is converted into
a mass density via a mass-to-light ratio $\Upsilon$, $\rho_*(\varpi,z)
= \Upsilon j_*(\varpi,z)$.  In most studies, $\Upsilon$ is assumed to
be constant over the entire galaxy.  Under this standard assumption,
our initial models were unable to produce an even remotely reasonable
fit to the kinematics in NGC 205. This is not surprising since
$\Upsilon$ is known to show significant radial variations in this
galaxy. For instance, Carter \& Sadler (1990), using simple King
models, find $\Upsilon_R = 9.4$ for the main body of the galaxy (with
an uncertainty of up to a factor of 2, especially at large radii) and
$\Upsilon_R = 2.35$ for the nucleus (equivalent to I-band values of
$\Upsilon_I = 6.95$ and $\Upsilon_I = 1.94$ respectively using
standard color corrections).  Allowing for a radial change in
$\Upsilon$ mandates a cautionary note: the problem of determining both
$\mh$ and $\Upsilon(r)$ is likely to be degenerate, and in principle a
variety of possible $\Upsilon$ profiles could undoubtedly be found
that fit the observational constraints equally well, each producing a
different best-fit (or upper limit) value for $\mh$.

We produced an independent estimate of $\Upsilon_I$ in NGC 205 by
using the 3-integral models to fit the luminosity distributions and
the kinematical data separately for the nucleus and the main body of
the galaxy.  The luminosity density of the nucleus is well
approximated by a Gaussian with FWHM $\sim$ 0\Sec26 (or $\sigma =
0\Sec11$) and the nucleus can therefore be regarded as the region
within $r \leq 0\Sec33$ (corresponding to $\sim 3 \sigma$ of the
Gaussian fit to the luminosity density of the nucleus). The main body
can be regarded as the region beyond $r \geq 0\Sec55$ ($\sim 5 \sigma$
of the Gaussian fit to the luminosity density of the nucleus) where
the color and luminosity profile change over to that of the main body
of the galaxy. Figure~\ref{fig:MbyL-In-Out} shows 1D curves of $\Delta
\chi^2 = \chi^2-\chi^2_{min}$ for the nucleus alone ($r \leq 0\Sec33$,
solid line) and for the outer region alone ($r \geq 0\Sec55$, red
dot-dash line) as a function of $\Upsilon_I$ for the case in which no
black hole is present. All models used a library with 8100 orbits.
The horizontal thin dot-dash line is at $\Delta \chi^2 = 1$. The solid
black line suggests that $(\Upsilon_I)_{nucleus} = 1.6\pm 0.8$
($1\sigma$ uncertainties) and $(\Upsilon_I)_{body} =
9\pm^{1.3}_{2.5}$, consistent within $1\sigma$ with the values derived
by Carter \& Sadler (1990).  We point out that neither estimate is
flawless. The adoption of King profiles is not motivated by current
data (or our current knowledge of the power law form of luminosity
profiles of elliptical galaxies).  Currently both methods assume that
it is possible to estimate M/L ratios separately for the two regions
by treating them as dynamically separate.  However in practice they
are a single dynamical entity and the M/L profile needs to be obtained
for both regions simultaneously. The most general method for finding
the radial profile$\Upsilon_I(r)$ would be to allow M/L to vary with
radius and obtain a non-parametric best-fit estimate for this function
for this using a method similar to the Schwarzschild method. This
beyond the scope of the current paper.

\begin{figure}
\figurenum{5}
\epsscale{0.75}
\plotone{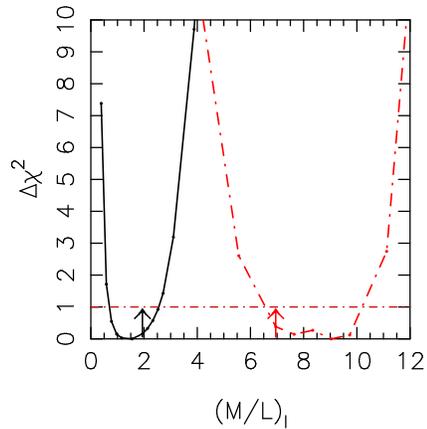}
\caption{1D curves of $\Delta \chi^2$ for the nuclear region ($r
\leq 0\Sec33$, solid line) and for the main body of the galaxy ($r
\geq 0\Sec6$, red dot-dash line) as a function of the I-Band M/L ratio
$\Upsilon_I$.  All models assume $\mh = 0$ and used 8100 orbits
each. The vertical arrows show the $\Upsilon_I$ values obtained by
Sadler and Carter (1990) using King-models.}
\label{fig:MbyL-In-Out}
\end{figure}

For lack of stronger constraints, we make the simplest possible
assumption about the radial variation of $\Upsilon$, namely we assume
$\Upsilon_I$ to be constant within 0\Sec33 (the nuclear region) and
beyond 0\Sec55 (the main body of the galaxy), and to vary linearly
between these two values at intermediate radii. Although the values of
$\Upsilon$ derived by Carter \& Sadler and those derived using the
Schwarzchild method are consistent, in the following section we will
carry out the dynamical modeling using both. In the first case (the
``Carter-Sadler profile'') we adopt $\Upsilon_I = 1.94$ for $r\leq
0\Sec33$, $\Upsilon_I = 6.95$ for $r \geq 0\Sec55$ and
linearly-varying between these two radii. In the second case (the
``Schwarzschild profile'') we adopt $\Upsilon_I = 1.6$ for $r\leq
0\Sec33$, $\Upsilon_I = 8.5$ for $r \geq 0\Sec55$ and linearly-varying
in between. In what follows we will indicate these ``primary''
profiles as $\Upsilon_I^*(r)$, and introduce, in the dynamical
modeling, a scaling factor, $S_\Upsilon =
\Upsilon_I(r)/\Upsilon_I^*(r)$, as a free parameter.  The second free
parameter is the mass $\mh$ of the central black hole. For libraries
which adopted the Carter-Sadler $\Upsilon$-profile we constructed
dynamical models for $17$ different values of $\mh$ in the range $0
\le\mh\le 10^7$\msun~ and $\sim 13$ values of $S_\Upsilon$. For
libraries which adopted the Schwarzschild $\Upsilon$-profile we
constructed dynamical models for $15$ different values of $\mh$ in
the range $0 \le\mh\le 10^6$ \msun~ and $13$ values of
$S_\Upsilon$.

\clearpage
\subsection{Regularization}
\label{sec:regularization}

It is customary to smooth, or ``regularize'', the orbital solutions,
either by including a penalty function with adjustable smoothing
parameter (e.g. VME04), by imposing a ``maximum entropy'' constraint
(e.g. Gebhardt et al. 2003) or by imposing a local smoothness
constraint in phase space (Cretton et al. 1999). As discussed at
length in VME04, the results can depend strongly on the value of the
smoothing parameter $\lambda$. An incorrectly chosen $\lambda$ can
generate misleading results; for instance, too great a degree of
smoothing has the same effect as limiting the range of allowed orbital
populations, and can give a spurious, best-fit value of $M_{BH}$ even
when no value is preferred by the data. Since the optimal choice of
$\lambda$ depends on the data set in question, we constructed models
for nine different values of $\lambda$ between $10^{-7}$ and $10$. In
the interest of time, all models used the Carter-Sadler
$\Upsilon$-profiles and libraries of ~4500 orbits.

\begin{figure}
\figurenum{6}
\epsscale{0.75}
\plotone{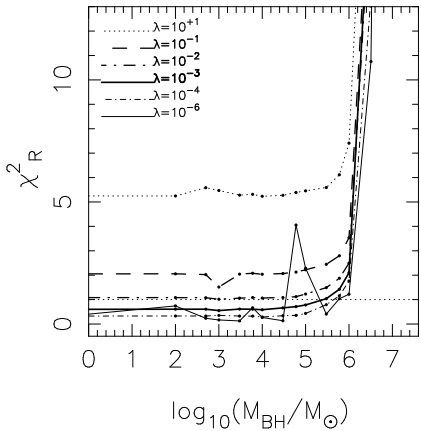}
\caption{1D reduced $\chi^2$ ($\chi^2_R$) curves as a function of
black hole mass ($\mh$) for 6 different values of the regularization
(smoothing) parameter $\lambda$. All models assume $\Upsilon_I=1.1$.
See text for further details.}
\label{fig:chivslam}
\end{figure}

1D reduced $\chi^2$ curves ($\chi^2_R$) for a fixed $S_\Upsilon = 1.1$
(the $\Upsilon$-scaling factor defined in \S~\ref{sec:mbylprofile})
are plotted in Figure~\ref{fig:chivslam} for six values of $\lambda$.
($\chi^2_R$ is computed for DOF=100 as discussed later in
\S~ref{sec:carter-sadler}). Most of the curves are characterized by a
nearly flat valley in $\chi^2$ at $\mh < 10^5$\msun.
Figure~\ref{fig:chivslam} also shows that for smallest value of
$\lambda=10^{-6}$ (thin solid curve) the $\chi^2_R$ plot is noisy and
oscillates randomly from one model to the next. For $\lambda = 10^{-4}
- 10^{-2}$ the curves are qualitatively similar: flat from $\mh = 0 -
10^5$\msun~ followed by a sharp upturn at higher $\mh$. For $\lambda =
10^{-4}$ the $\chi^2_R = 0.23$ in the flat region, for $10^{-3}$ the
$\chi^2_R = 0.62$ in the flat region (thick solid line) and for
$\lambda = 10^{-2}$ the $\chi^2_R =1.07$. For values of $\lambda$
larger than $\lambda = 10^{-2}$ $\chi^2_R$ increases above one, and
the upper limit on $\mh$ depends more sensitively on $\lambda$, with
$\mh$ decreasing as $\lambda$ increases. We could choose either
$\lambda = 10^{-3}$ or $\lambda = 10^{-2}$. We chose $\lambda =
10^{-3}$ as the value that gives the largest $\chi^2_R <1$. In the
remainder of this paper (unless otherwise noted) this is our preferred
value of $\lambda = 10^{-3}$.  Other, ostensibly, ``objective''
methods to determine the optimal value of the regularization parameter
exist but they are either not well motivated physically or extremely
costly computationally.
In the interest of limiting the number of parameters varied we use of
$\lambda= 10^{-3}$ also for models using libraries of 8100 orbits. Our
quoted results of the upper limit on $\mh$ and choice of $S_\Upsilon$
are all for this value of $\lambda$, but this does not imply that
other values of $\lambda$ give statistically less probable results.

\section{Results of 3-Integral modeling}
\label{sec:results}

\subsection{The Zeroth-Order Model}
\label{sec:carter-sadler}

We begin our analysis by discussing the results of the dynamical
models under the following conditions: 1) all models use a 4500 orbit
library; 2) the regularization parameter $\lambda$ is set equal to
$10^{-3}$; 3) the $\Upsilon_I^*$ profile is that given by Carter and
Sadler (1990) (\S 3.1); and 4) the STIS kinematical information is extracted
using Template 1 (\S 2.1). In the next subsection we will change some
of these conditions and investigate how the results are affected.

Figure~\ref{fig:bend-stisGH} shows 2-dimensional contour plots of the
total $\chi^2$ as a function of $S_{\Upsilon} =
\Upsilon_I(r)/\Upsilon_I^*(r)$ and $\mh$. The models covered 17 values
of $\mh$ and 13 values of $S_{\Upsilon}$ (indicated by the grid of
points); the contours are plotted at $\Delta(\chi^2) = 2.30, 4.61,
6.17, 9.21, 11.8, 18.4$ which correspond to 68.3\%, 90\%, 95.4\%,
99\%, 99.73\%, 99.99\% confidence intervals assuming two degrees of
freedom (DOF) (but see note below).  Beyond the 6th contour the spacing
of contours is arbitrary.  Two sets of contours are shown, depending
on which set of kinematical constraints are fitted. While every model
is required to fit the luminosity (mass) constraints (for a total of
304 constraints), the dashed (red) contours correspond to models which
were required to fit the ground-based kinematical data only (a total
of 62 constraints). The solid (black) contours show the results when
the fit is also required to be constrained by the STIS data ($V$,
$\sigma$, $h_3$ and $h_4$) for an additional 40 constraints. 


\begin{figure}
\figurenum{7}
\epsscale{0.75}
\plotone{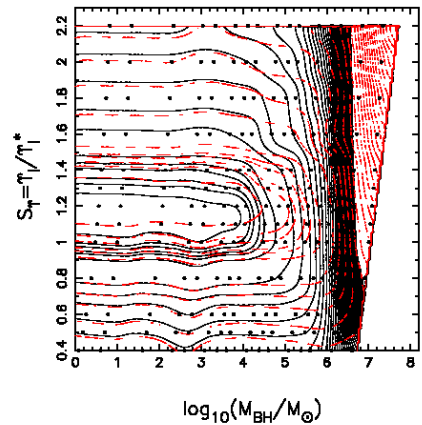}
\caption{ Contours of constant $\chi^2$ derived from fits of the
data to 3-Integral axisymmetric models.  Dashed contours: fits to the
Bender et al. (1991) ground-based data.  Solid contours: fit to the
combined ground-based $+$ STIS GH moments derived using Template
1. Each dot represents a set of parameters ($\mh, S_\Upsilon$) for
which a 3-Integral dynamical model was constructed. The total number
of orbits used in each case was $N_o=4500$. The first 3 contours of
each type are at $1\sigma, 2\sigma, 3\sigma$ confidence intervals
respectively (68.5\%, 90\% and 95\%). The regularization parameter was
$\lambda = 10^{-3}$.}
\label{fig:bend-stisGH}
\end{figure}

Both sets of $\chi^2$ contours are striking in their lack of
dependence on $S_{\Upsilon}$, which is determined from these plots to
be $1.13\pm^{0.08}_{0.1}$ ($1\sigma$ uncertainty).  Although a minimum
in the $\chi^2$ contours is not seen, the addition of the small-radius
spectral data from STIS allows us to reduce the upper limit on $\mh$
{\it by an order of magnitude}, as shown more clearly in
Figure~\ref{fig:Big-Small}. The curves labeled {\bf A, B} are 1D cuts
through Figure~\ref{fig:bend-stisGH} at the minimum in $S_{\Upsilon}=
1.1$: curve {\bf A} is a fit to the ground and STIS data and gives an
upper limit of $\mh \approx 8\times10^3$\msun~ while curve {\bf B} is
the fit to the ground-based data alone and gives an upper limit of
$\mh \approx 10^5$\msun~ (both upper limits are given at the 1$\sigma$
confidence level, shown by the the horizontal dotted lines drawn above
each $\chi^2$ curve).  Note that the 1D $\chi^2$ plots shown in
Figure~\ref{fig:Big-Small} {\em are not marginalized} with respect to
$S_\Upsilon$, but are merely cuts through the 2-parameter $\chi^2$
plots. (Curves {\bf C} and {\bf D} of Fig\ref{fig:Big-Small} are
described in the following section.)

\begin{figure}
\figurenum{8}
\epsscale{0.75}
\plotone{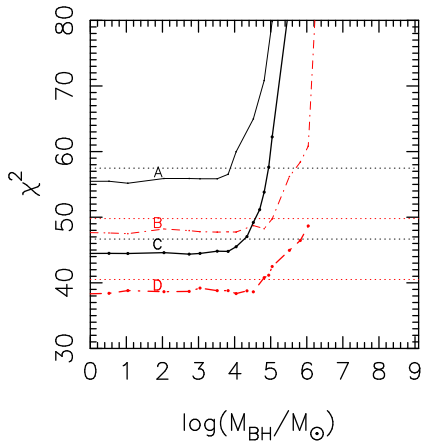}
\caption{1D $\chi^2$ curves which illustrate the dependence of the
upper limit on $\mh$ on orbit library size and on smoothing parameter
$\lambda$. Individual curves are described in the text.}
\label{fig:Big-Small}
\end{figure}
A well known assumption of the use of the $\chi^2$ statistic is that
the problem is linear in the parameters being estimated. An additional
unstated assumption in all discussions of confidence interval
estimation (e.g. Press et al. 1992) is that the number of fitted
parameters is less than the number of data points -- in other words,
that the parameter estimation problem is over-constrained by the
data.  It is only in this case that quantities like ``number of
degrees of freedom'' (DOF=$N_{\rm data} - N_{\rm parameters}$) make
any sense. But in Schwarzschild modeling, the number of parameters
(orbits + potential parameters) is typically far greater than the
number of data points, i.e., the problem is under-constrained by the
data.Indeed, one typically finds that a variety of choices for the
parameters ($\mh, \Upsilon$) can reproduce the data equally well
(VME04), and if it were not for positivity constraints on the orbital
occupation numbers, this degeneracy would be even greater. We know of
no discussion in the statistical literature that deals adequately with
this situation, and for this reason, the standard prescriptions for
estimating confidence regions (e.g. Press et al.) should be considered
suspect when applied to Schwarzschild modelling.

It is therefore necessary to for us to first clarify how the
confidence intervals are computed.  The quality of a particular fit
was assessed by comparing the total $\chi^2$ value of the fit to the
number of degrees of freedom (DOF), as being ``the number of
independent data points - number of {\it potential } parameters''.
Thus the number of parameters is assumed to be 2 (while actually it is
much larger due to the orbits). For the NGC 205 dataset, the number of
independent data points is 102 (for the STIS plus ground-based
kinematical data) or 62 (for the ground-based kinematical data only;
note that the luminosity constraints are ``model constraints'' and not
``data constraints''), while the number of parameters is two
($S_{\Upsilon}$ and $\mh$), giving DOF=100 (STIS+ground based data) or
DOF=60 (ground-based data only). This DOF was used in
\S~\ref{sec:regularization} to compute reduced-$\chi^2$ ($\chi^2_R$)
which is expected to be $\sim 1$ for a good fit. It has become
customary in this field to further assume that the confidence
intervals on the parameters being estimated (namely $\mh$ and
$\Upsilon$) can be obtained from $\Delta(\chi^2)$ distribution with
the number of degrees of freedom = to the number of parameters.

 Note also that unless otherwise stated the
upper limits quoted are $1\sigma$ confidence limits. In a few
instances we also quote $3\sigma$ limits. Also we caution that the
contour plots are {\em not} reliable for determining upperlimits since
they are plotted with an algorithm which uses adaptive 2D regression -
we always compute upperlimits from 1D cuts through the $\chi^2$ plots.
It is also worth noting that when judging whether, for instance, a
model with a black hole is preferred over a black-hole-free model, we
urge the skeptical reader to follow our practice of closely inspecting
the kinematical profiles predicted by the different models and
deciding whether either model is significantly better at reproducing
the data in the region where the black hole's influence would be felt.

\subsection {Robustness of the Results}

In this section we explore the dependence of the results obtained
above on 1) the size of the orbit library; 2) the choice of the
smoothing parameter $\lambda$; 3) the stellar template used in
recovering the kinematical data from the STIS spectra; and 4) the
choice of $\Upsilon$ profile.  The results of all models discussed in
this section are summarized in Table 3.
\begin{table}
  \caption{Results of the Dynamical modeling}
  \begin{tabular}{lcccclcl}
  \hline
Run & No. & $\lambda$ & $\Upsilon_I$\tablenotemark{1} & Template & Constraints\tablenotemark{2}& $S_{\Upsilon}$ & $\mh$ \\
No. & Orbits &  &  &  & &  & (\msun) \\ \hline
1   & 4500  &   $10^{-3}$  &  CS  &  1  &   G     & $1.13\pm^{0.08}_{0.1}$  &  $<1\times10^5$    \\
2   & 4500  &   $10^{-3}$  &  CS  &  1  &   G-Sgh & $1.13\pm^{0.08}_{0.1}$  &  $<8\times10^3$    \\
3   & 8100  &   $10^{-3}$  &  CS  &  1  &   G     &  fixed at 1.1   &  $<6.4\times10^4$    \\
4   & 8100  &   $10^{-3}$  &  CS  &  1  &   G-Sgh &  fixed at 1.1   &  $<2.2\times10^4$    \\
5   & 8100  &   $10^{-4}$  &  CS  &  1  &   G-Sgh &  fixed at 1.1   &  $<3.5\times10^4$    \\
6   & 8100  &   $10^{-2}$  &  CS  &  1  &   G-Sgh &  fixed at 1.1   &  $<1.1\times10^4$    \\
7   & 8100  &   $10^{-1}$  &  CS  &  1  &   G-Sgh &  fixed at 1.1   &  $<0.8\times10^4$    \\
8   & 4500  &   $10^{-3}$  &  CS  &  2  &   G-Sgh &  $1.13\pm^{0.08}_{0.1}$  &  $<8\times10^3$    \\
9   & 8100  &   $10^{-3}$  &  CS  &  1  &   G-Slos-Sgh & fixed at 1.1   &  $<2.2\times10^4$    \\
10  & 8100  &   $10^{-3}$  &  CS  &  2  &   G-Slos-Sgh & fixed at 1.1   &  $<2.2\times10^4$    \\
11  & 4500  &   $10^{-3}$  &  S   &  1  &   G-Sgh & $0.95\pm^{0.05}_{0.08}$  &  $<1.0\times10^4$    \\
12  & 8100  &   $10^{-3}$  &  S   &  1  &   G-Sgh & fixed at 1.0 & $<1.3\times10^4$    \\
\hline
\tablenotetext{1}{Carter-Sadler $\Upsilon$ profile (CS) or Schwarzschild $\Upsilon$ profile (S)}
\tablenotetext{2}{Ground-based data only (G); Ground+STIS GH-moments (G-Sgh); Ground+STIS Central LOSVD +STIS GH for other apertures (G-Slos-Sgh)}
\end{tabular}
\end{table}

We consider first the size of the orbit library. Curves {\bf C, D} in
Figure~\ref{fig:Big-Small} are analogous to curves {\bf A, B} except
they were derived from models which used 8100 orbits. In the interest
of time, these models were run at $\Upsilon_I(r)/\Upsilon_I^*(r)=1.1$,
for 17 different values of $\mh$. It is evident from the figure that
increasing the size of the orbit library relaxes the upper limit on
$\mh$, at least when both the STIS and ground based data are fitted.
Using the ground based-data only, we now derive an upper limit of $\mh
= 6.4\times10^4$\msun~ (curve {\bf D}), while adding the STIS
constraints produces an upper limit $\mh \lap 2.2\times10^4$\msun~
(curve {\bf C}) (at $1\sigma$ confidence) and an upper limit of $\mh
\lap 3.9\times10^4$\msun~ (at $3\sigma$ confidence). This sensitivity
to the size of the orbit library was discussed in VME04, where we
advocated using the largest orbit library that was computationally
feasible to obtain the full range of statistically allowed
solutions. (We do not believe that the decrease in upper limit of
$\mh$ derived from the ground-based data is statistically
significant. An examination of the contribution to $\chi^2$ shows that
this number is sensitive to the the noise in the large radius data
which has no bearing on the mass of a central black hole.)

As mentioned earlier, the choice of $\lambda$ depends on the size of
the orbit library. For the 4500 orbit models, $\lambda=10^{-3}$ was
identified in \S 3.2 as the best value based on the fact that this
gave a $\chi^2_R$ closest to, but below one. The effect of
changing $\lambda$ for this larger orbit library is shown Table~3
(Run No. 4-7). When the full dataset (STIS+ground based) is fitted with a
smaller smoothing parameter $\lambda=10^{-4}$, the $1\sigma$ upperlimit
increases by 60\% to $\mh \lap  3.5\times10^4$\msun~ (with $\chi^2_R
\sim 0.23$. When the same dataset is fitted with a larger smoothing
parameter $\lambda=10^{-2}$, the upperlimit deccreases by 50\% to $\mh
\lap  1.1\times10^4$\msun~ (with $\chi^2_R \sim 0.9$. And when the
dataset is fitted with $\lambda=10^{-1}$, the upperlimit decreases
even further to $\mh \lap  0.77\times10^4$\msun~ (with reduced
$\chi^2 \sim 1.4$.

As $\lambda$ increases, the fit to the data becomes smoother and as a
consequence $\chi^2$ also increases. We know of no truly robust way to
pick smoothing such that it does not strongly influence the derived
estimate of $\mh$. In the absence of a robust choice of $\lambda$ we
simply quote the value obtained for $\lambda=10^{-3}$ for reasons
given in \S~\ref{sec:regularization}.

\begin{figure}
\figurenum{9}
\epsscale{0.75}
\plotone{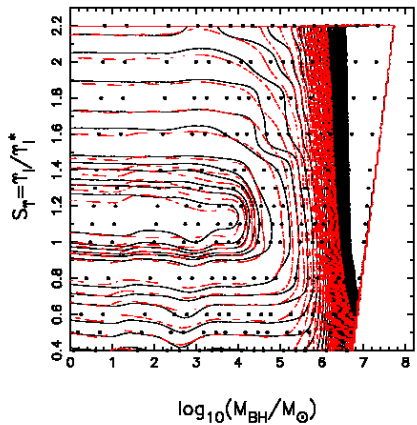}
\caption{Comparison of $\chi^2$ contours for models that fit the
STIS GH moments obtained using the two different stellar templates.
Solid (black) contours are for models that fit kinematics derived using
Template 1 and dot-dash (red) contours are for models that fit
kinematics from Template 2.}
\label{fig:stisT1-T2}
\end{figure}

The effect of using a different template in extracting the kinematical
constraints from the STIS data (\S 2.1) is shown in
Figure~\ref{fig:stisT1-T2}.  This figure is directly comparable to
Figure~\ref{fig:bend-stisGH}, since it was produced using a 4500 orbit
library and $\lambda = 10^{-3}$.  As expected based on the consistency
between the $V, \sigma, h_3, h_4$ values shown in
Figure~\ref{fig:STIS-kin}, the fits to the kinematics using the two
stellar templates give virtually identical limits on both $S_\Upsilon$
and $\mh$.

\begin{figure}
\figurenum{10}
\epsscale{0.75}
\plotone{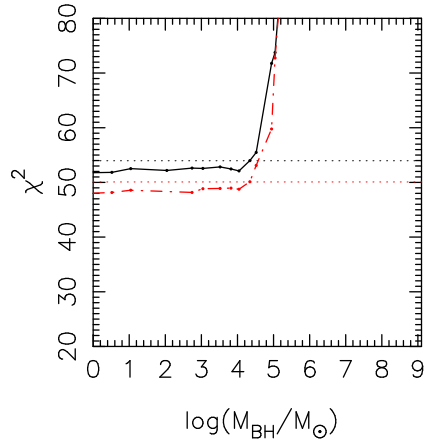}
\caption{1D-$\chi^2$ curves obtained by fitting the central LOSVD
and the Gauss-Hermite moments for other apertures. The solid line is
from fitting the central LOSVD and GH moments for other apertures
obtained with Template 1 and the dot-dash line is from fitting the
kinematics obtained with Template 2.}
\label{fig:chi1D-LOS1-LOS2-big}
\end{figure}

However, as discussed in \S 2.1 and shown in
Figure~\ref{fig:Central-LOSVD}, the two different templates do produce
significantly different shapes for the LOSVD in the innermost
apertures. While these differences are disconcerting, it is
nevertheless desirable to try to include the extra information
contained within the central LOSVD, since it is here that the effect
of a central black hole would make its presence most strongly felt
(e.g.  van der Marel 1994). Earlier authors (e.g. Gebhardt et al.
2000b) have modeled even strongly asymmetric central LOSVDs in the
detection of a nuclear black hole with axisymmetric models. These
authors attribute the asymmetry in the central LOSVD to an
off-centered central aperture as well as due to the presence of
obscuring dust.

Figure~\ref{fig:chi1D-LOS1-LOS2-big} shows two 1D-$\chi^2$ plots
obtained by fitting the full LOSVD (sampled at $\Delta V = 20$\kms)
for the central STIS aperture, and the Gauss-Hermite moments for other
apertures.  The solid and dot-dashed lines refer to fits to the
kinematics derived using Template 1 and Template 2 respectively; as
before, the two horizontal dot-dash lines represent the $1\sigma$
confidence limit above the minimum of the corresponding curve. All
models used orbit libraries with 8100 orbits and a smoothing parameter
$\lambda = 10^{-3}$. Although the models that fit kinematics from
Template 1 (solid line black) give a systematically higher $\chi^2$
because of the difficulty in fitting the highly asymmetric central
LOSVD, both LOSVDs produce identical upper limits on $\mh$ of
$2.2\times 10^4$\msun~ - which is also identical to that obtained from
fitting the GH moments for all the apertures (curve {\bf C} in
Fig~\ref{fig:bend-stisGH}). The insensitivity of the results to
whether the full LOSVD is fitted or not, is likely to be a result of
the fact that the sphere of influence of a putative central black hole
in NGC 205 is not resolved, and therefore the number of high velocity
stars which are able to affect the high velocity wings of the LOSVD is
proportionately small.  While the LOSVD from Template 1 might point to
a detection of heavy wings, the increased difficulty of fitting the
asymmetry with an axisymmetric model does not appear to result in an
increase in the upper limit.

\begin{figure*}
\figurenum{11}
\epsscale{0.8}
\plotone{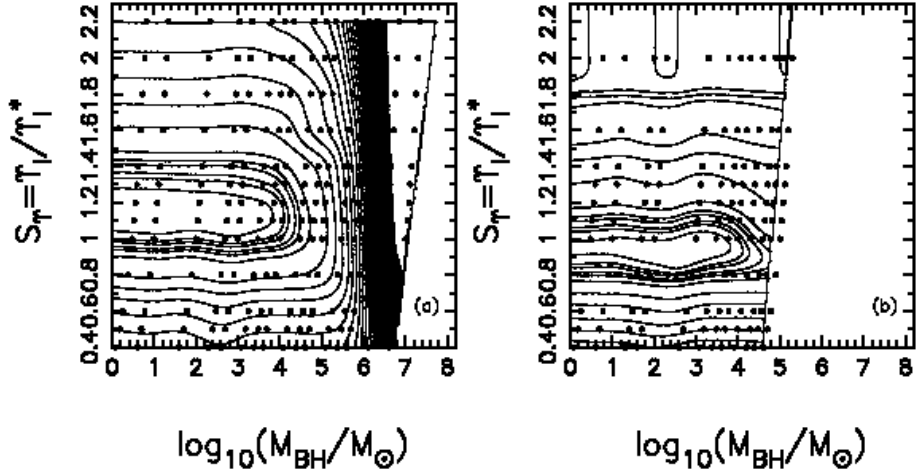}
\caption{Comparion of $\chi^2$ contour plots obtained by fitting
data with libraries (of 4500 orbits) constructed with the two
different $\Upsilon$ profiles: (a) for orbit libraries that used the
Carter-Sadler profile; (b) for libraries constructed with the
Schwarzschild profile. }
\label{fig:Mbyl-2D-prof}
\end{figure*}


We conclude this section by discussing how the estimate of $\mh$ is
affected by choice of $\Upsilon$ profile. We would have preferred to
do this for the models with 8100 orbits; however because of the
modeling time involved, we first ran a full set of models with the
smaller (4500) orbit library, varying both $S_\Upsilon$ and $\mh$, and
then ran a subset of the larger, more time-intensive models at the
value of $S_\Upsilon$ corresponding to the minimum found from the
previous set of models. Models using 4500 orbits were therefore
constructed using the Schwarzschild $\Upsilon$ profile for 14
different values of $\mh$ from $0.1$ to $10^5$\msun~ and 13 values of
$S_\Upsilon$.  The models were then fitted to the luminosity (mass)
constraints, plus the GH moments obtained using Template 1.  A
two-dimensional $\chi^2$ plot for this set of models is shown is
Figure~\ref{fig:Mbyl-2D-prof}(b); the two-dimensional $\chi^2$ plot
with models that used the Carter-Sadler profile is reproduced again in
Figure~\ref{fig:Mbyl-2D-prof}(a) for comparison. As before, the three
inner most contours represent the $1\sigma, 2\sigma$ and $3\sigma$
($\Delta\chi^2 = 2.3, 4.6, 6.1$ for DOF=2) confidence intervals
respectively.  In Figure~\ref{fig:chi1D-S-ups} we plot cuts through
Figures~\ref{fig:Mbyl-2D-prof}(a) (solid curve) and
Figures~\ref{fig:Mbyl-2D-prof}(b) (dot-dash curve) for a fixed
$\mh=10^2$. The Schwarzschild $\Upsilon$ profile models show a minimum
at $S_\Upsilon = 0.95\pm^{0.05}_{0.08}$
(Fig.~\ref{fig:Mbyl-2D-prof}(b)) (as opposed to $S_\Upsilon =
1.13\pm^{0.08}_{0.1}$ in the case of the Carter-Sadler profile
(Fig.~\ref{fig:Mbyl-2D-prof}(a)). The intercepts with the horizontal
dot-dash line indicate the $1\sigma$ confidence intervals. (Note that
a few additional models were run in each case to fully explore the
minium). Figure~\ref{fig:chi1D-S-ups} shows that the best fit values
of $S_\Upsilon$ for the two models are significantly different at the
$1\sigma$ level. Values of $S_\Upsilon$ close to the minima in
Fig~\ref{fig:chi1D-S-ups} were then chosen to construct a subset of
models with 8100 orbits for all values of $\mh$. The 1D $\chi^2$
profiles resulting from these larger orbit libraries are shown in
Figure~\ref{fig:Mbyl-1D-prof} as a dot-dashed curve (the results using
the Carter-Sadler profile are shown as a solid curve for comparison).
The Schwarzchild profile gives $\mh < 1.3\times 10^4$\msun, compared
to $\mh < 2.2\times 10^4$\msun~ for the Carter-Sadler profile.

Finally, we note that the overall $\chi^2$ produced by the best fit
models for the two $\Upsilon$ profiles are virtually indistinguishable
from each other (the minima differ by $\Delta \chi^2 \sim 1$), and
therefore there is no reason to prefer one profile over the other.
This is an indication that the 3-integral modeling problem is
inherently degenerate as discussed at length in VME04.

\begin{figure}
\figurenum{12}
\epsscale{0.75}
\plotone{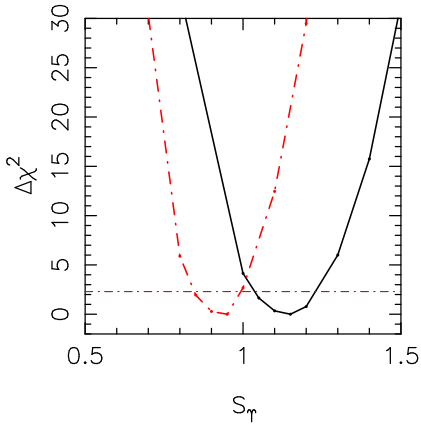}
\caption{Cuts through 2D$\chi^2$ contour plots in
Fig~\ref{fig:Mbyl-2D-prof} at $\mh = 10^2$\msun~ for the Carter-Sadler
profile (solid black curve); for the Schwarzschild profile (dot-dash
red curve).}
\label{fig:chi1D-S-ups}
\end{figure}

From the discussion above we conclude that $2.2\times 10^4$\msun~
represents a firm upper limit to the mass of the central black hole in
NGC 205 (Table 3).  Figure~\ref{fig:kinematics-4MBH} shows the model
fits to the kinematical data for $S_{\Upsilon} = 1.1$ (for models
obtained with the Carter-Sadler profile and 8100 orbits per library)
and four different black hole masses: no black hole (dot-dash line),
$\mh = 1.1\times10^3$\msun~(dashed line); $\mh = 2.2\times 10^4$\msun~
(solid line, corresponding to the 1$\sigma$ confidence upper limits
for models with 8100 orbits and $\lambda = 10^{-3}$), $\mh = 5\times
10^4$ \msun~(dotted line, corresponding to the 5$\sigma$ limit for the
same model). The plot confirms that the two smaller values of $\mh$
produce statistically equivalent fits to the data, and that the
central STIS velocity dispersion, $\sigma \sim 21$\kms, provides the
strongest constraint on the upper limit on the mass of a massive
central black hole.

\begin{figure}
\figurenum{13}
\epsscale{0.75}
\plotone{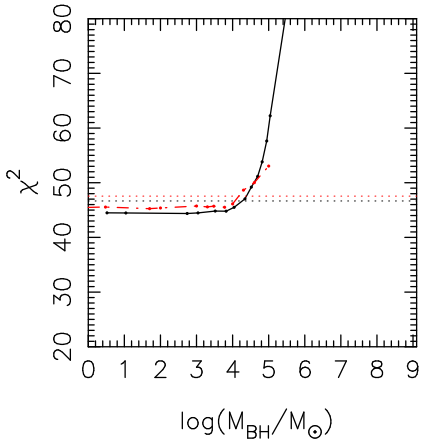}
\caption{Comparion of 1D $\chi^2$ curves at fixed values of
$S_\Upsilon$ obtained by fitting data with libraries of 8100 orbits
for libraries that used the Carter-Sadler profile (solid curve); for
libraries constructed with the Schwarzschild profile (dot-dash
curve).}
\label{fig:Mbyl-1D-prof}
\end{figure}

\section{Implications for Black Hole Scaling Relations}
\label{sec:scaling}

While the firm dynamical detection of an IBH has yet to be achieved in
any galaxy, there are now two secure upper limits, in M33 (Merritt et
al. 2001; Gebhardt et al. 2001) and in NGC 205 (this paper).  Here we
investigate the implications of these upper limits for the form of the
scaling relations that have been established using SBH detections in
more massive galaxies.  The tightest of these relations is the $\ms$
relation as derived from central aperture dispersions (Ferrarese \&
Merritt 2000).  This relation is plotted in Figure~\ref{fig:msigma},
as updated by Ferrarese \& Ford (2004), using only those galaxies (25
in number) with secure dynamical detections.  The solid line in
Figure~\ref{fig:msigma} shows the best fit regression line:
${M_{BH,8}} = (1.66 \pm 0.24) {\sigma_{200}}^{4.86 \pm 0.43}$
($M_{BH,8}\equiv \mh/10^8$\msun, $\sigma_{200}\equiv\sigma/200$ km
s$^{-1}$).  Also plotted are the upper limits on $\mh$ in M33 and NGC
205.  For M33, we conservatively adopt $\mh<3000$ \msun~ (Merritt
\etal 2001), rather than the smaller value of $1000$ \msun~ claimed by
Gebhardt et al. (2001).

\begin{figure*}
\figurenum{14}
\epsscale{0.8}
\plotone{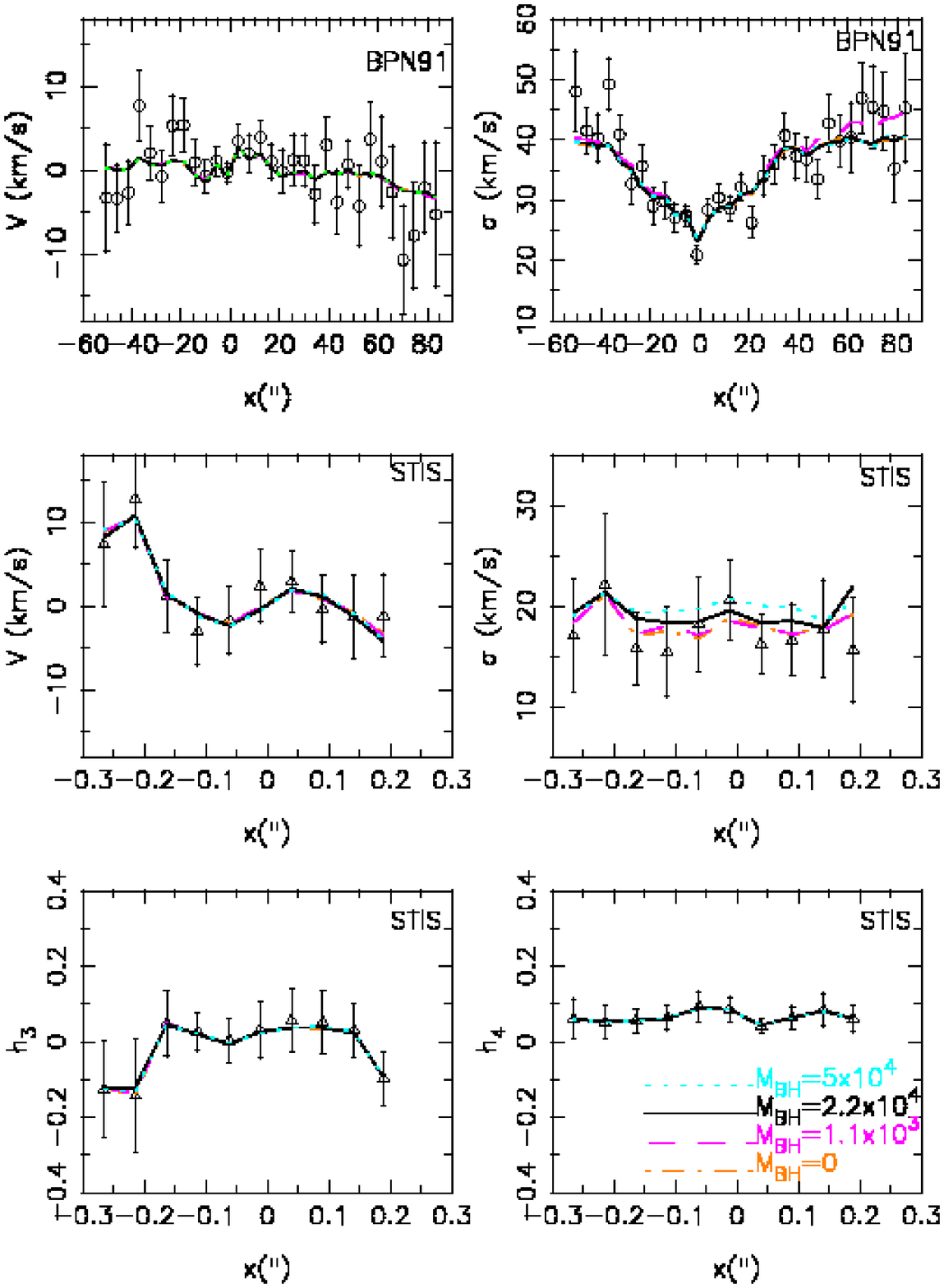}
\caption{Fit to ground based data of Bender et al. (1991)
(top two panels) and STIS data (bottom 4 panels) with libraries of
8100 orbits for 4 different value of $\mh$ as indicated by the labels. The
solid line is the fit for the upper limit $\mh = 2.2\times 10^4$\msun.}
\label{fig:kinematics-4MBH}
\end{figure*}

Also shown (dotted line in Figure~\ref{fig:msigma}) is the relation
claimed by Tremaine et al. (2002), using a velocity dispersion
measured by those authors within one effective radius.  The difference
in slopes between the two relations is barely significant in a
statistical sense, but becomes critical when the relations are
extrapolated to the low mass regime. The upper limits on the IBH
masses in M33 and NGC 205 are both inconsistent with the shallower
$\ms$ relation advocated by Tremaine et al. (2002) but not with the
Ferrarese \& Ford (2004) or Ferrarese \& Merritt (2000) relations.

Given the importance of determining whether these upper limits on
$\mh$ are consistent with the scaling relations determined at higher
masses, we must address the issue of the slope.  Among the differences
in the way that the two groups construct the $\ms$ relation, probably
the most important is sample definition.  Gebhardt et al. (2000a) and
Tremaine et al.(2002) adopt a less restrictive criterion than Ferrarese
\& Merritt (2000) in establishing the reliability of a SBH detection;
in particular, they include several detection based on data which do
not resolve the SBH sphere of influence (comprising roughly $30\%$ of
the claimed SBH detections based on stellar kinematics), and others
(e.g. NGC 3379; Gebhardt et al. 2000b) for which the authors
themselves acknowledge that a model with no SBH fits the data
precisely as well as a model containing a SBH. Tremaine et al. (2002)
also include SBH mass estimates which are deemed by their own authors
to have systematic uncertainties that exceed the quoted errors
(NGC1068, Greenhill et al. 1996; NGC 4459, NGC 4596, Sarzi et
al. 2001; NGC224, Bacon et al. 2001).  Empirically, the scatter in all
SBH scaling relations is seen to decrease, sometimes dramatically,
when detections based on galaxies which do not resolve the sphere of
influence are removed from the sample (Merritt \& Ferrarese 2001;
Ferrarese \& Ford 2004; Graham et al. 2001; Marconi \& Hunt 2003).
The slope is also found to increase when such detections are removed
from the sample (Ferrarese \& Ford 2004).


\begin{figure}
\figurenum{15}
\epsscale{0.95}
\plotone{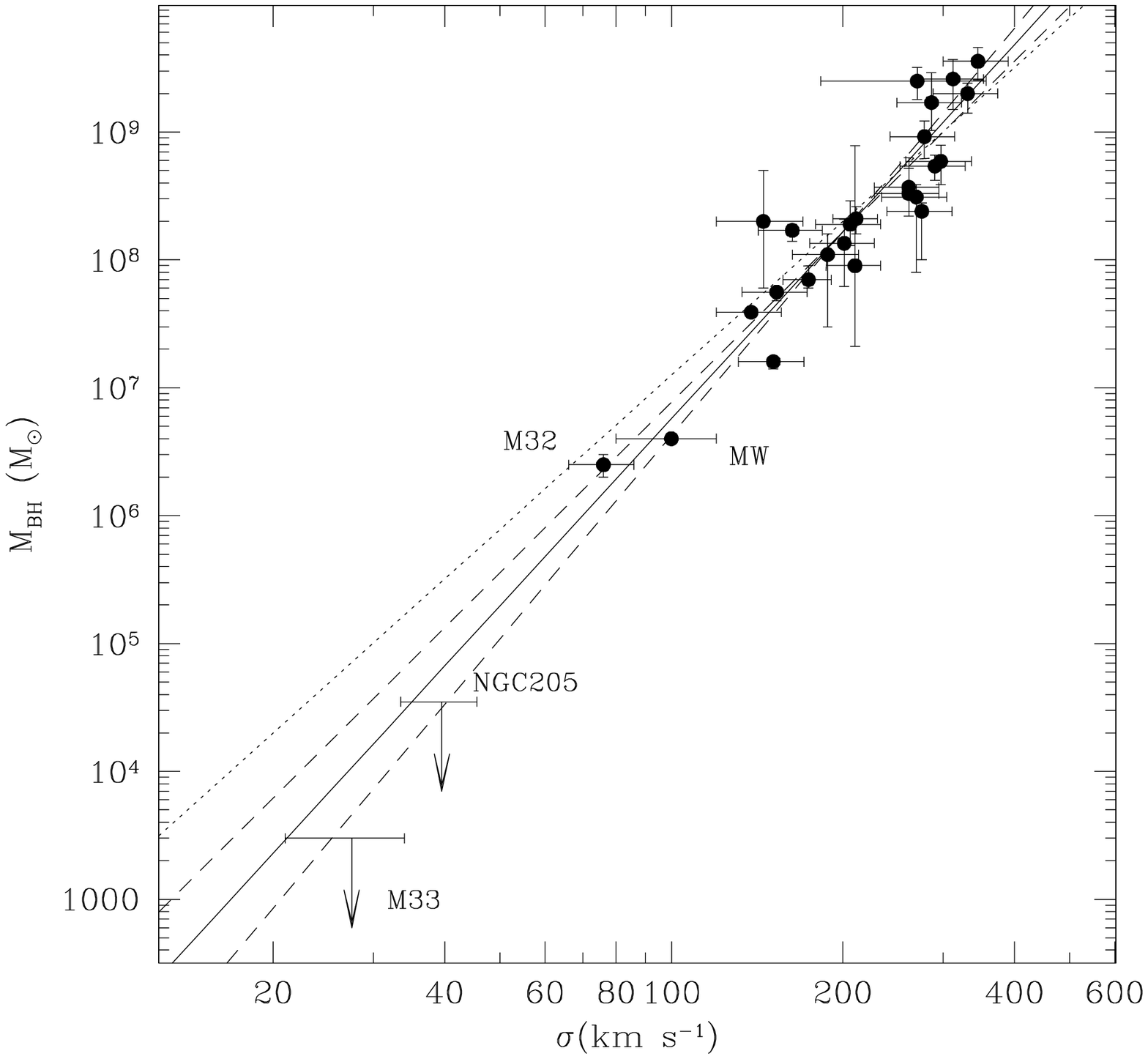}
\caption{$\ms$ relation. Solid line is the best fit from Ferrarese \&
Ford (2004), with 1$\sigma$ errors on the slope shown by the dashed
line. The dotted line is the best fit from Tremaine et~al. (2002).
Only masses based on data which resolve the sphere of influence are
plotted; the stellar velocity dispersion $\sigma$ is as defined in
Ferrarese \& Merritt (2000).}
\label{fig:msigma}
\end{figure}

Our conclusion is therefore that a steeper slope to the $\ms$
relation, such as that derived by Ferrarese \& Merritt (2000) and
Ferrarese \& Ford (2004), is more appropriate. Therefore, the upper
limits on the masses of nuclear IBH in M33 and NGC 205 are not
inconsistent with the extrapolation of the $\ms$ relation to the
low-mass regime. Put another way, while there is still no compelling
dynamical evidence for IBHs in these galaxies, we can not yet rule out
the presence of IBHs with masses similar to those predicted by the
established scaling relations.

Rather than examine whether the upper limits on $M_{BH}$ in M33 and
NGC 205 are consistent with the $M_{BH}-\sigma$ relation, we can
rederive that relation under the {\it assumption} that it applies to
both galaxies, using the upper limits on $M_{BH}$ as constraints.  We
apply a technique called ``regression with censored data'' that is now
standard amongst statisticians and has been applied in a few
astronomical contexts (e.g. Isobe, Feigelson \& Nelson 1986). Two
censored regression methods are in wide use: the EM algorithm
(Dempster, Laird \& Rubin 1977), a maximum likelihood method that
assumes a normal distribution of the ordinate values about the
best-fit line; and a modification of the EM algorithm by Buckley \&
James (1979), hereafter the BJ algorithm, which infers the
$Y$-distribution from the measurements themselves and is believed to
be more robust to non-normal errors.  Unfortunately, neither algorithm
takes into account measurement errors in either
variable,\footnote{This was also the case for the regression algorithm
used by Gebhardt et al. (2000a) in their derivation of the slope of
the $\mh-\sigma$ relation.}  and for neither is there a well-accepted
scheme for estimating the uncertainties in the fitted coefficients.
In addition, these algorithms assume that the upper limits are
precisely known.

Table 4 gives coefficients in the fit of the data in Figure 14 to
\begin{equation}
\log_{10} M_{BH,8} = \alpha\log_{10}\sigma_{200} + \beta.
\end{equation}
We used the implementations of the EM and BJ algorithms provided by
Isobe et al. (1986).  The first line in Table 4 gives the slope
($4.52\pm 0.36$) and intercept when M33 and NGC 205 are excluded.  The
slope is slightly lower than the value ($4.86$) quoted above; the
latter was derived from a regression algorithm that accounts for
measurement errors (in general, ignoring measurement errors leads to
spuriously low estimates of the slope).  The remaining lines give the
fitting parameters from the EM and BJ algorithms, respectively, under
two assumptions about the upper limits on $M_{BH}$ in M33 and NGC 205.
Including the two upper limits always increases the inferred slope, to
values in the range $5.2\lap\alpha\lap 5.7$.  Had we been able to
include measurement errors in the censored regression algorithms,
these slopes would probably have been even greater.  We consider these
values to be the best current estimates of the slope of the
$M_{BH}-\sigma$ relation, under the assumption that {\it  the relation
extends to spheroids as faint as M33 and NGC 205}.

\begin{table}
  \caption{Fits to $\log M_{BH,8}=\alpha\log\sigma_{200} + \beta$}
  \begin{tabular}{cccc}
  \hline
$\mh(M33)$ &  $\mh(NGC 205)$ & $\alpha$ & $\beta$  \\
 \hline
 --- & --- &  $4.52\pm 0.36$ &  $0.22\pm0.06$\\
$<3000$ & $<10,000$ & $5.48$($5.43$) & $0.17$($0.17$) \\ 
$<1000$ & $<10,000$ & $5.67$($5.58$) & $0.16$($0.17$) \\ 
$<3000$ & $<35,000$ & $5.29$($5.23$) & $0.18$($0.18$) \\ 
$<1000$ & $<35,000$ & $5.52$($5.43$) & $0.17$($0.17$) \\ 
\hline
\end{tabular}
\end{table}

\section{Morphological Constraints on the Presence of a Massive BH in NGC 205}
\label{sec:morphology}

The presence or absence of a massive black hole in NGC 205 might be inferred
indirectly, from the observed structure of the nucleus combined with
evolutionary arguments.  We begin by contrasting the structure of NGC
205 with that of the other well-resolved Local Group spheroids, in
M31, M32, M33, and the Milky Way (Lauer et al. 1992, 1998; Genzel et
al. 2003).  Each of these galaxies exhibits a steep luminosity
profile, $\rho\sim r^{-\gamma}$, $1.5\lap\gamma\lap 2$, inward of
$\sim 1$ pc, with densities at $0.1$ pc that range from $\sim
10^6M_\odot$ pc$^{-3}$ (M33) to $\sim 10^7M_\odot$ pc$^{-3}$ (M32).
Only M33 and NGC 205 exhibit a core, with radius $\sim 0.2$ pc in both
galaxies.  The inferred central density of M33, assuming $M/L_v=0.4$
in solar units, is $\sim 2\times 10^6M_\odot$ pc$^{-3}$ (Lauer et
al. 1998), compared with $\sim 3\times 10^5M_\odot$ pc$^{-3}$ in our
mass model of NGC 205.  M33 and NGC 205 also have similar central
kinematics: the 1D central rms velocity is $\sim 20-30$ km s$^{-1}$ in
both galaxies, compared with much higher values at the centers of the
other Local Group galaxies.

The lack of a dynamical detection of a massive BH in NGC 205 also
implies an upper limit to its gravitational influence radius, $r_h
\lap 0.16 {\rm pc}\ M_{BH,4} \sigma_{20}^{-2}$ ($M_{BH,4}\equiv
M_{BH}/10^44M_\odot$, $\sigma_{20}\equiv \sigma/20$ km s$^{-1}$).
While no clear morphological signature is apparent at $r_h$ in any of
the other Local Group galaxies, even those belived to contain massive
BHs, the power-law density cusps in M32 and the Milky Way are at least
consistent with the $\rho\sim r^{-7/4}$ density profile that is
established around a BH on a relaxation time scale $T_r$ (Bahcall \&
Wolf 1976), where
\begin{equation}
T_r = {0.34\sigma^2\over G^2m_\star \rho \ln\Lambda}
\approx 1.4\times 10^8 {\rm yr}\sigma_{20}^3 \rho_5^{-1} \left(\ln\Lambda_{10}\right)^{-1}
\end{equation}
($\rho_5\equiv\rho/10^5M_\odot$ pc$^{-3}$,
$\ln\Lambda_{10}\equiv\ln\Lambda/10$.)  The outer radius of such a
cusp is expected to be $\sim 0.2r_h$ (Preto, Merritt \& Spurzem 2004),
making it unobservable in NGC 205 even if we adopt our upper limit on
$M_{BH}$.

Figure~\ref{fig:relax} plots $T_r$ as a function of radius in NGC 205;
we also plot the time scale $T_{coll}$ for star-star collsions,
\begin{equation}
T_{coll} = \left[16\sqrt{\pi}n\sigma r_\star^2\left(1+\Theta\right)\right]^{-1}
\approx 8.5\times 10^{10} {\rm yr}\sigma_{20}^{-1} n_{5}^{-1} 
\Theta_{200}^{-1}
\end{equation}
where $n_5$ is the stellar number density in units of $10^5$ stars
($M_\odot$) per $pc^3$, $m_\star$ and $r_\star$ are a typical stellar
mass and radius respectively, and $\Theta\equiv
Gm_\star/(2\sigma^2r_\star)$.  We evaluted these two time scales as
functions of radius using kinematical quantities taken from our
Schwarzschild solutions, setting
$\sigma^2=(\sigma_x^2+\sigma_y^2+\sigma_z^2)/3$, $m_\star=M_\odot$,
$r_\star=R_\odot$, and $\ln\Lambda=10$.  The results were found to be
almost the same whether the assumed BH mass was $\sim 0$ or $2.2\times
10^4$~\msun; Figure~\ref{fig:relax} shows the results for $M_{BH}=0$.
While the stellar collision time is not directly relevant to the
issues addressed in this Section, we note that $T_{coll}$ is
sufficiently long even at $0.1$ pc in NGC 205 that a typical
solar-type star is unlikely to have suffered a collision in its
lifetime.  In this respect NGC 205 is similar to other Local Group
spheroids, all of which have $T_{coll}(0.1{\rm pc})$ of order
$10^{11}$ yr.  Hence we ignore collsions in what follows.  Collisions
might nevertheless have played a role in establishing the $M/L$
gradient in NGC 205, a subject that will be discussed in more detail
in a later paper.

Figure~\ref{fig:relax} shows that the relaxation time in NGC 205 is
quite short, with a central value of $\sim$ a few $\times 10^7$ yr.
Among the Local Group spheroids, only M33 has a comparably short
central relaxation time, $T_r\approx 5\times 10^6$ yr.  The low value
of $T_r$ in M33 prompted the suggestion (Hernquist, Hut \& Kormendy
1991) that the nucleus of this galaxy might have undergone core
collapse.  In the remainder of this Section, we discuss whether a
similar case can be made for NGC 205, and whether the changes induced
in the nuclear morphology of NGC 205 by core collapse would depend on
the presence of a massive BH.

The case for core collapse in M33 is based on its short central
relaxation time.  If core collapse has {\it not} already occurred in
M33, it would be expected to take place in a time $T_{cc}\approx
10^2T_r(0)\approx 3\times 10^8$ yr, with $T_r(0)$ the current central
relaxation time.  Unless we live at a special time, it is therefore
very likely that the M33 nucleus has already undergone core collapse.
The constant-density core of M33 might be a result of the
binary-driven re-expansion that takes place following collapse.

This argument is not quite as strong in the case of NGC 205.  We note
that the exact constant of proportionality between $T_{cc}$ and
$T_r(0)$ depends on the details of the stellar distribution function,
and can vary from $\sim 10$ to $\sim 10^3$ (Quinlan 1996).  However
the nuclear density profile of NGC 205 is reasonably close to the
self-similar, $\rho\propto r^{-2.2}$ form of late core collapse, and
the current central relaxation time is short enough that considerable
evolution toward core collapse must have already occurred.  In these
circumstances, the constant of proportionality should be close to its
asymptotic value, $T_{cc}\approx 330 T_r(0)$ (Spitzer 1987), and the
current central value of $T_r$ implies that the time remaining to core
collapse (if it has not already occured) is $\sim 1.6\times 10^{10}$
yr.  This time could be shortened somewhat if the central density
rises inward of the resolution limit, or if the nucleus contains a
range of stellar masses (e.g. G\"urkan et al. 2004).
However $T_{cc}$ is not short enough that we can conclude, as in M33,
that core collapse has definitely occurred, and this makes it more
difficult to draw definite conclusions about the effects of a massive
BH on nuclear evolution.

\begin{figure}
\figurenum{16}
\epsscale{.85}
\plotone{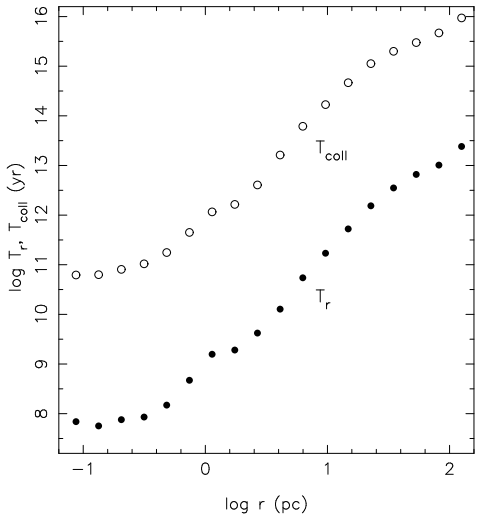}
\caption{Relaxation time $T_r$ and stellar collision time
$T_{coll}$ as functions of radius along the major axis in NGC 205, in
the model solution with $M_{BH}=0$.  See text for details.}
\label{fig:relax}
\end{figure}

If a massive BH were present at the center of NGC 205, how would this
affect the evolution toward core collapse or its subsequent effects on
nuclear morphology?  The theory of core collapse in the presence of a
BH has not been as widely developed as the theory of core collapse in
a purely stellar system, and most discussions make particular
assumptions about the efficiency of accretion of gas from
tidally-disrupted stars and its effect on the growth of the BH
(e.g. Shapiro 1977; Murphy, Cohn \& Durisen 1991; Baumgardt et
al. 2004).  In its early stages, core collapse around a BH is driven
by the same evaporation of high-velocity stars that drives classical
core collapse (Henon 1961).  Collapse is eventually halted, either by
binary formation, or by the input of heat due to capture or disruption
of stars by the BH.  The latter process is expeced to dominate the
former at densities like those of the nucleus of NGC 205 (Murphy, Cohn
\& Durisen 1991); the core then re-expands, on a time scale of order
the stellar consumption time, or $\sim T_r(r_h)$.  Outside of the BH's
sphere of influence, the density profile in a post-core-collapse
nucleus is probably similar to that in a nucleus without a BH
(e.g. Fig. 8b of Murphy et al. 1991).  Significant differences in the
rate of core collapse or its effects on the structure of the nucleus
would only be expected if the BH were large enough that stellar
capture or disruption dominated from the outset (Shapiro 1977).

We conclude that the structure and kinematics of NGC 205 do not allow
us to make a definitive statement about whether or not there is
morphological evidence for a central SBH. Core collapse may have
occurred, but it is also possible that the core is still evolving
toward collapse.

\section{Conclusions}
\label{sec:conclusions}

We discussed new HST ACS/HRC images and STIS spectra for the nuclear
region of NGC~205, a nucleated dwarf elliptical companion of the
Andromeda galaxy. The surface brightness profile derived from the
$I-$band images was combined with large scale ground based data from
Lee (1996) and deprojected non-parametrically to derive the 3-D
luminosity density. The kinematical information (velocity, velocity
dispersion and the GH $h_3$ and $h_4$ coefficients), extracted from
the STIS spectra by means of a maximum penalized likeliwood algorithm,
does not seem to depend significantly on the choice of stellar
template. The STIS spectra are complemented at large radii with data
from Bender et al. (1991). The full suite of data provides both
luminosity and kinematic information up to 1.4 arcmin from the center
(305 pc at the galaxy distance of 740 kpc).

In an effort to constrain the presence of a central black hole,
state-of-the-art 3-integral dynamical models based on orbital
superposition were constructed for a large number of the parameters
[$\mh,S_{\Upsilon}$], where $S_{\Upsilon} =
\Upsilon_I(r)/\Upsilon_I^*(r)$ controls the scaling of the radially
variable stellar M/L ratio relative to a ``nominal" value.

Several sets of models were computed and compared with the data, under
different assumptions regarding the number of orbits, the value of the
regularization parameter, and the radial dependence of the
mass-to-light ratio.  Model predictions were compared with two sets of
observed kinematics, derived from the STIS data using different
template spectra. We also compared the constraints imposed with and
without the inclusion of the central LOSVD.

Under no set of assumptions were we able to recover a best-fit value
of $M_{BH}$ in NGC 205.  The upper limit which we derive on $M_{BH}$
is somewhat dependent on the assumptions made in the modelling, as
expected based on earlier work (VME04).  We found upper limits ranging
from $\sim 0.8\times 10^3$\msun to $\sim 3.5\times 10^4$\msun.  Even
for the largest of these upper limits, the radius of influence of the
black hole would not be resolved by our data, hence the failure of the
modelling to produce a firm detection is not surprising. An upper
limit of $2.2\times 10^4$\msun for the black hole in NGC 205 -- our
preferred value (\S4.2) -- is consistent with the extrapolation to the
low-mass regime of the $M_{BH}-\sigma$ relation derived from black
holes more massive than $\sim 10^6$\msun.

We wish to thank the anonymous referee for the many suggestions which
significantly helped in improving this manuscript. MV thanks Eric
Emsellem for detailed comments on an earlier version of this paper and
M.Y. Poon for assistance with extracting published ground based data
from the literature.  This work was supported by STScI grant GO-09448
and grants AST-0206031, AST-0420920 and AST-0437519 from the NSF and
grant NNG04GJ48G from NASA.



\end{document}